\documentclass[a4paper,fleqn,usenatbib]{mnras}

\usepackage{newtxtext,newtxmath}

\usepackage[utf8]{inputenc}
\usepackage[T1]{fontenc}
\usepackage{ae,aecompl}

\usepackage{graphicx}	
\usepackage{amsmath}	
\usepackage{amssymb}	

\usepackage{graphicx}

\usepackage{bm}
\usepackage{tabularx}
\usepackage[roman]{parnotes}
\usepackage[normalem]{ulem}

\usepackage{siunitx}
\usepackage{color}

\newcommand{\Msun}{\text{M}_{\odot}}

\input{mathdefs.sty}

\DeclareMathOperator{\erf}{erf}
\DeclareMathOperator{\erfc}{erfc}

\DeclareMathOperator*{\diag}{diag}

\newcommand{\crdchart}{\phi_x}
\newcommand{\surfchart}{\phi_f}
\newcommand{\volf}[1]{\text{vol}_{#1}} 
\newcommand{\data}{\mathcal{D}}
\renewcommand{\vct}[1]{\bm{#1}}
\renewcommand{\mat}[1]{\mathbfss{#1}}

\newcommand{\ortho}{\nu} 
\newcommand{\para}{\tau} 

\newcommand{\cvm}{\bm{\Sigma}}
\newcommand{\intcvm}{\bm{\Sigma}_{\text{int}}}

\newcommand{\relpars}{\vct{\theta}}
\newcommand{\intpars}{\vct{\varphi}}

\newcommand{\intdist}{p_\text{int}}
\newcommand{\postdist}{p_\text{post}}

\newcommand{\detdist}{p_\text{det}}

\newcommand{\ellE}{\mathcal{E}}

\newcommand{\tnote}[1]{$^{#1}$}

\newcommand{\linmixerr}{\texttt{linmix\_err}}
\newcommand{\fitexy}{\texttt{FITEXY}}

\newcommand{\msigma}{$M_{\text{BH}}$--$\sigma$}

\newcommand{\add}[1]{#1}
\newcommand{\st}[1]{}

\title[Non-linear correlations with intrinsic scatter]{A geometric approach to
non-linear correlations with intrinsic scatter}

\author[P. Pihajoki]{
Pauli Pihajoki$^{1}$\thanks{E-mail: pauli.pihajoki@iki.fi}
\\
$^{1}$ University of Helsinki, Department of Physics, Gustaf Hällströmin katu 2a, 00560
Helsinki, Finland
\\
}

\date{Accepted XXX. Received YYY; in original form ZZZ}

\pubyear{2017}

\begin{document}
\label{firstpage}
\pagerange{\pageref{firstpage}--\pageref{lastpage}}
\maketitle

\begin{abstract}
    We propose a new mathematical model for $n-k$-dimensional non-linear
    correlations with intrinsic scatter in $n$-dimensional data.  The
    model is based on Riemannian geometry, and is naturally
    \add{symmetric with respect to the measured variables and} invariant
    under coordinate transformations. We combine the model with a
    Bayesian approach for estimating the parameters of the correlation
    relation and the intrinsic scatter.
    \add{A side benefit of the approach is that censored and truncated
    datasets and independent, arbitrary measurement errors can be
    incorporated.}
    We also derive analytic likelihoods for the typical astrophysical
    use case of linear relations in $n$-dimensional Euclidean space. We
    pay particular attention to the case of linear regression in two
    dimensions, and compare our results to existing methods.
    Finally, we apply our methodology to the well-known \msigma{}
    correlation between the mass of a supermassive black hole in the
    centre of a galactic bulge and the corresponding bulge velocity
    dispersion. The main result of our analysis is that the most likely
    slope of this correlation is $\sim 6$ for the datasets used, rather
    than the values in the range $\sim4\text{--}5$ typically quoted in
    the literature for these data.
\end{abstract}




\section{Introduction}

An important question in all the sciences is whether different measured
observables seem to be connected by some form of mathematical relation,
or whether they seem to be completely independent. If a relation is
suspected, it then becomes important to estimate the type of relation
connecting the measurements and to estimate its parameters.

Historically, these problems have been most often approached by the use
of various correlation coefficients and linear and non-linear regression.
Linear regression in particular has been the subject of lively debates and much
research over the past two centuries or so, ever since first derived in
the guise of least-squares regression by Gauss (or possibly Legendre,
see \citealt{stigler1981}). The discussion has included such points of
contention as how to treat the variables symmetrically -- that is how to
avoid dividing the variables into dependent and independent variables --
and whether this is necessary
\citep{pearson1901,boggs1987,isobe1990,feigelson1992,robotham2015}.
\add{
This issue is particularly important when investigating the hypothesis
that the observables obey some mathematical relation. By definition,
such a relation as a geometric structure (e.g.\ a line, a plane, or some
more complicated subset) is unique, even though it may have many
equivalent formulations. As such, any method used for estimating such a
relation should eventually yield the same structure even if the
variables are permuted or a different variable is chosen as the
dependent variable. \emph{This requires that the method must involve all
observed variables in an essentially symmetric way.}
However, this is not true for least squares regression and many other
similar approaches as well (see Section~\ref{sc:comparison}). This fact
has been well known for a long time \citep[see
e.g.][]{pearson1901,isobe1990}, but it appears somewhat
underappreciated. Since the focus of this paper is on formulating and
estimating relations between observables, we will pay close attention to
this symmetry invariance throughout.
}

Another recurrent theme is the question of how to treat data subject to
censoring (i.e.\ lower and upper limits), truncation (non-detections)
and heteroscedastic, independent and correlated errors in general, and
finally, how to incorporate intrinsic scatter, or intrinsic uncertainty
in the regression hyperplane (see e.g.\ \citealt{kelly2007},
\citealt{hogg2010} and \citealt{robotham2015} and the references
therein).
\add{
Intrinsic scatter can be intuitively understood to mean that the data
probability distribution can be non-zero also in the neighbourhood of
the subset defined by the relation, that is \emph{we also allow for data that may
be somewhat off the relation} and thus not exactly satisfying the
equation(s) defining the relation. 
For example, in
the case of a line in two dimensions, intrinsic scatter would manifest
as a data distribution that is not a linear delta function ridge, but a
`fuzzy' line instead. Despite this intuitive obviousness,
intrinsic scatter has a history of various somewhat non-rigorous
definitions, such as through an additional additive component in
measurement errors. In this paper, we will present a more rigorous
definition of intrinsic scatter which is also usable in non-Euclidean
contexts.}

Specific examples \add{of papers addressing some of the problems above}
include \citet{pearson1901}, who introduced
a least squares method to fit lines and hyperplanes based on minimizing
residuals orthogonal to the regression plane (OR, orthogonal
regression). Later, algorithms were developed to solve this problem for
non-linear relations as well (see e.g.\ \citealt{boggs1987} and
\citealt{boggs1988} and the references therein). In \citet{kelly2007} and
\citet{hogg2010}, problems related to outliers, truncation, censoring
and intrinsic scatter are solved with a fully Bayesian approach, with
some limitations. Namely,
\citet{kelly2007} requires specifying a single dependent variable,
and it is for this variable only that censoring is supported, while
\citet{hogg2010} only consider the two-dimensional case.
Finally, \citet{robotham2015} extends the results in \citet{hogg2010} to
\add{$n$ dimensions, for arbitrary $n$}, and presents analytic likelihoods for
$n-1$-dimensional hyperplanes with intrinsic scatter (i.e.\ lines in a
plane, planes in three-dimensional space and so on),
\st{in $n$-dimensional} \add{assuming}
data with Gaussian errors.

However, there still exist some remaining issues related to fitting non-linear
relations to data that have been discussed to a much lesser extent.
One of these is the notion of pre-existing geometry in the measured
quantities, such as for angular quantities, or measurements of points on
curved surfaces (however, see e.g.\ \citealt{pennec2006} and
\citealt{calin2014}). 
Another is the question of a proper characterisation of
intrinsic scatter for non-linear relations, and how to incorporate
intrinsic scatter when the $n$-dimensional data is not well described by an
$n-1$-dimensional subspace (i.e.\ of codimension one), but an
$n-k$-dimensional subspace (codimension $k$), for an arbitrary $k>1$.
A suitable resolution of these issues is of interest, since it would enable
powerful hypothesis testing via finding the most likely value of $k$ and
distribution of intrinsic scatter for each proposed linear
or non-linear relation simultaneously.

In this paper, we propose a solution to these issues by formulating the
concept of non-linear relations with intrinsic scatter of general
codimension $k$ through Riemannian geometry. The novelty of our approach
is in extending the idea of intrinsic scatter to curved spaces and
correlations that are non-linear and have codimension greater than one,
that is, not restricted to linear hyperplanes of dimension $n-1$. The
approach also accommodates arbitrary measurement errors, censoring and
truncation. Furthermore, we
give analytic results for the likelihood and posterior probability for
linear $n-k$-dimensional relations in Euclidean spaces. These are
easy to implement in fitting codes, and useful for hypothesis testing by
enabling quick determination of the most likely codimension of a potential
correlation in $n$-dimensional data, along with the parameters of the
intrinsic scatter.

In Section~\ref{sc:relations}, we give definitions of correlation
relations and intrinsic scatter distributions through the use of
Riemannian submanifolds and an intrinsic coordinate system. The
definitions allow the concept of intrinsic scatter distributions to be
smoothly extended to non-linear relations and correlations of
codimension greater than one. We also give a Bayesian solution to the
problem of finding the most likely parameters for a given relation and
intrinsic scatter, incorporating censoring, truncation and independent,
general distributions of measurement errors. In
Section~\ref{sc:specialcases} we apply our formulation to the special
case of linear relations in Euclidean spaces. In particular, we extend
the results in \citet{robotham2015} to $n-k$-dimensional subspaces with
intrinsic scatter, 
\st{a result we believe to be both novel and highly useful.}
\add{which correspondingly extends their usefulness to a much wider
range of applications.}
We also extensively discuss the implications of the
formalism in the case of linear regression in two dimensions, with
comparisons to well-established existing methods. Finally, in
Section~\ref{sc:msigma}, we apply our methodology to the well-known
\msigma{} relation between the mass $M_\text{BH}$ of a supermassive black hole in a
galactic bulge and the bulge velocity dispersion $\sigma$. 
\add{Finally,} we \st{sum up} \add{present a summary of} the
paper in Section~\ref{sc:summary}.

We note that Section~\ref{sc:relations} is by nature
somewhat technical, and relies heavily on concepts in Riemannian
geometry. However, the salient points of the formulation we propose are
illustrated in Figures \ref{fig:surfchart} through
\ref{fig:2d-dist-examples}. For the reader with immediate applications
in mind, we suggest looking at the likelihood functions derived for the
special cases of an $n-1$-dimensional linear relation, equation~\eqref{eq:ndimllh}, 
$n-k$-dimensional linear relation, equation~\eqref{eq:nk-likelihood} and
the results for the two-dimensional case, equations
\eqref{eq:llh-single} and \eqref{eq:upperx}--\eqref{eq:loguppery}.

\section{Relations and intrinsic scatter}\label{sc:relations}

\subsection{Relations as submanifolds}

The dual aim of this paper is firstly to give a geometric formulation of
relations with intrinsic scatter that is intuitive and adapted to the geometry
of the relation and the space of measurements. Secondly, 
to fit these relations to data, we aim to provide a method that
is fully Bayesian and places no observable in a privileged
position, i.e.\ there is no division into independent and
dependent variables. In addition, we wish to accommodate truncated and censored
datasets and independent and general distributions of measurement errors.
To this end, we need to carefully define what we mean by a relation and
its intrinsic scatter. We also need equal care in defining what we mean by being
off the relation to be able to define intrinsic scatter in the first place.

In this paper, we will consider relations as submanifolds of the
manifold of all possible combinations of measurement values, 
which we take to have a
Riemannian geometry. Being off the relation will be related to the
geodesic (shortest path) distance between a point in the measurement
manifold, representing a single possible combination of measurements, and the
relation submanifold.
In this approach, the joint distribution of the measured observables is the
primary object of interest. It is this joint distribution that we wish to model
with the relation submanifold and the intrinsic scatter. Consequently, the
distributions of the individual observables, obtained as marginal distributions,
are of secondary interest only.

We start by defining a relation as a smooth map \mbox{$f:M\fromto\fR^k$}, where $M$ is
an $n$-dimensional manifold with a Riemannian metric $g_{ab}$ and $k\leq
n$. The relation $f$ is fully specified by the $k$ component functions
$f_i$, so that $f(m)=(f_1(m),\ldots,f_k(m))$, for $m\in M$.
\add{Furthermore, the relation $f$ may depend on $n_p$ parameters
$\vct{\relpars}\in\fR^{n_p}$.}
In the following, we assume
that we are always working with some local coordinate chart
$\crdchart:M\fromto\fR^n$ and at times use a shorthand $\vct{x}$
for the point $m\in M$ for which
$\crdchart(m) = (x^1,\ldots,x^n) = \vct{x}$. 
These coordinates represent the
observable quantities (e.g.\ angle, charge, distance, velocity, mass and so on) in
the chosen units, or as scale-free logarithmic measurements.
The manifold $M$ then represents all possible combinations of measured values
for these observables.

Given $M$ and $f$, the level set 
\begin{equation}
    \begin{split}
        S &= \{m\in M | f(m) = \vct{0}\} \\
          &= \bigcap_{i=1,\ldots,k}S_i =\bigcap_{i=1,\ldots,k}\{m\in M | f_i(m) = 0\}
    \end{split}
\end{equation}
defines an $n-k$-dimensional (or of codimension $k$) subset $S\subseteq M$.%
\footnote{%
    A general level set would be $f(m) = \vct{c}$, $\vct{c}\in\fR^k$, but we
    assume that the constant $\vct{c}$ has been subsumed in the
    definition of $f$.
}
We further require that $S$ is a regular level set, that is, there are no $m\in M$ for
which $f(m)=0$ and the pushforward $\ud f_m$ fails to be surjective,
so that $S$ is additionally an embedded (or regular) submanifold of $M$,
as is each $S_i$ \citep[see][for a complete discussion]{lee2013}.
$S$ then represents the locus of our relation. 

\begin{figure*}
    \begin{tabular}{cc}
        \includegraphics[width=0.5\textwidth]{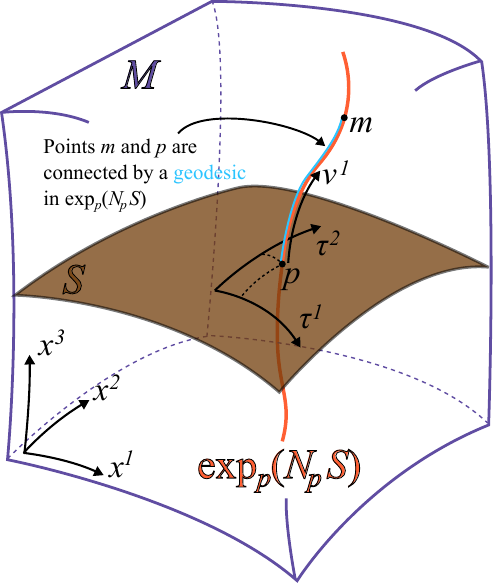} & 
        \includegraphics[width=0.5\textwidth]{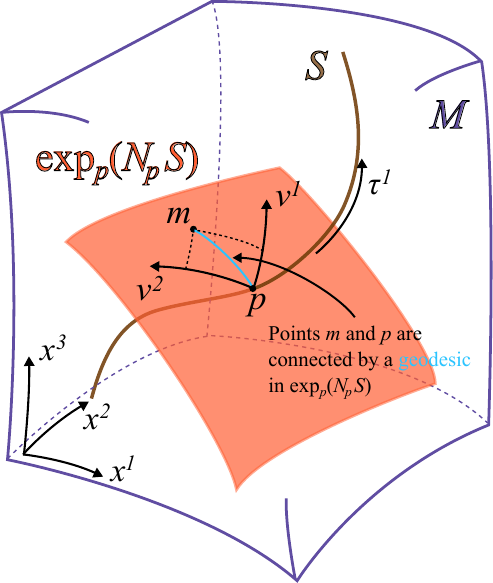}
    \end{tabular}
    \caption{
        A figure demonstrating the relationships between the measurement manifold
        $M$ (shown in violet), the manifold $S$ defined by the relation
        (brown), the exponential mapped normal space $\exp_p(N_p S)$ (orange)
        and the coordinate charts $\crdchart$ and $\surfchart$ 
        in a case when $\dim(M)=3$. Shown are cases where $\dim(S)=2$
        (left) and $\dim(S)=1$ (right). The coordinates 
        for a point $p$ on $S$ are either $\crdchart(p)=(x^1(p), x^2(p), x^3(p))$
        or $\surfchart(p) = (\para^1(p), \para^2(p), 0)$ (left) or
        $\surfchart(p) = (\para^1(p), 0, 0)$ (right). For $m$,
        $\surfchart(m)=(\para^1(m), \para^2(m), \ortho^1(m))$ (left) or 
        $\surfchart(m)=(\para^1(m), \ortho^1(m), \ortho^2(m))$ (right). The
        origin and orientation of the components $\vct{\para}$ can be set arbitrarily
        on $S$, but the origin of the $\vct{\ortho}$ is fixed on the
        point $p$.
    }
    \label{fig:surfchart}
\end{figure*}

The most geometrically obvious way to define being off the
relation is to relate it to the \emph{geodesic distance of a point from the
submanifold $S$.} To this end, it is useful to construct a new coordinate system
defined with the help of the normal bundle
\begin{equation}
        NS = \bigcup_{p\in S} N_p S 
           = \bigcup_{p\in S} \{z\in T_pM | g(z,w) = 0, \forall w\in T_pS\} 
\end{equation}
of $S$, essentially consisting of all the tangent vectors of $M$ that are
orthogonal to $S$ at each point $p$ on $S$.
We now define a coordinate
system $\surfchart$ on $M$ by defining
$\surfchart(m) = (\vct{\para}(p), \vct{\ortho}(z)) \in
\fR^{n-k}\times\fR^{k}$ 
where 
$m = \exp_p(z)$, $(p,z)\in NS$, and
$\vct{\para}$ is an arbitrary coordinate
system on $S$ and $\vct{\ortho}$ is an orthonormal coordinate system on 
$N_pS$.
Here $\exp_p: TM\fromto M$ is the exponential mapping
around point $p$. In these coordinates, the geodesic distance $\delta$ between
$m$ and the submanifold $S$ is just $\delta = \sqrt{g(z,z)}$.
Intuitively, the coordinates $\surfchart$ correspond to covering
the original manifold $M$ using normal coordinates around each point $p\in S$,
or $M\subseteq \cup_{p\in S}\{p\}\times EN_p$, where we have used a shorthand $EN_p=\exp_p(N_pS)$.
See the explanatory Figure~\ref{fig:surfchart}. The spaces $EN_p$ contain
all the points that can be reached from $p$ by geodesics orthogonal to $S$ at
$p$.  However, the subspaces $EN_p$ do not necessarily fit together
neatly.
In general, $\surfchart^{-1}$ fails to be injective. This happens when two or more geodesics
$\gamma_i(t)=\exp_{p_i}(t z_i)$, $i\in I\subset\fZ_{+}$, intersect each other at
some $m\in M$. Usually normal coordinates are only defined up to these points, 
which define the boundary of the region where
$\exp_p$ is a diffeomorphism.
In this case, however, it is advantageous to let $EN_p$ contain the points
reachable from $p$ by geodesics of unlimited length. In this case, 
$\surfchart$ may be
multivalued, such that $\surfchart(m) =
\{(\vct{\para}(p_i),\vct{\ortho}(v_i)\}$, $i\in I$.
This is discussed in the Sections~\ref{sc:intdist} and~\ref{sc:intscat}.
When $M$ is complete and connected and $S$ is closed, it is
known that $\surfchart^{-1}$ is surjective \citep{wolf1996}.
In some pathological cases $\surfchart^{-1}$ may fail to be surjective,
such as when the relation submanifold $S$ has sharp corners, but in this paper
we consider only cases where $S$ is sufficiently smooth.

\subsubsection{A concrete example}

\begin{figure}
    \includegraphics[width=0.5\textwidth]{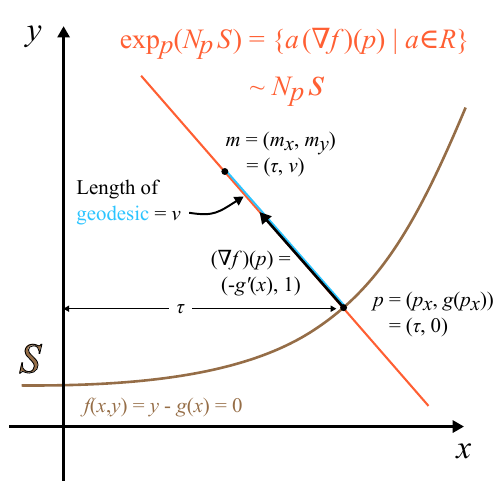}
    \caption{An example of a one-dimensional relation submanifold $S$ in a
    2-dimensional Euclidean manifold $M$. The coordinate transformation between
    $\crdchart$ and $\surfchart$ can be solved with equations
    \eqref{eq:2d-crdtrans}.}
    \label{fig:2d-example}
\end{figure}

A prototypical astrophysical
case is that of a relation between two real-valued observables. In this case,
the manifold $M$ is $\fR^2$, the local
coordinates are the identity map $\crdchart(x,y) = (x,y)$ 
and the relation is $f(x,y;\vct{\relpars})$, parameterized by $n_p$
parameters $\vct{\relpars}\in\fR^{n_p}$.
The normal spaces are $N_{p}S = \{\lambda\cdot(\nabla f)(p) |\lambda\in\fR \}$ and we can identify
$\exp_{p}(N_{p}S)$ with $N_{p}S$ itself. If we can put our
relation $f$ in the form $y-g(x)=0$, we can find the coordinate
transformation $\surfchart\circ\crdchart^{-1}$ as follows. If $m\in M$
is the point under consideration, and $p\in S$ is the point on $S$ geodesically
closest to $m$, with $\crdchart(m) = (m_x,m_y)$ and $\crdchart(p)=(p_x,
g(p_x))$, then we can take for example $\para(p_x,g(p_x)) = p_x$. With this
definition, we can in principle solve $\ortho$ and $\para$ as a function
of $m_x$ and $m_y$ from
\begin{subequations}
\label{eq:2d-crdtrans}
\begin{align}
    m_x &= p_x + \ortho \frac{-g'(p_x)}{\sqrt{1+g'(p_x)^2}} \\
    m_y &= g(p_x) + \ortho \frac{1}{\sqrt{1+g'(p_x)^2}}.
\end{align}
\end{subequations}
See Figure~\ref{fig:2d-example}.
Unfortunately, analytic solutions for these equations are not
straightforward to derive except in the case where $g(x)$ is linear.

\subsection{Intrinsic distributions}\label{sc:intdist}

We now have a coordinate system $\surfchart$ with which we can define
intrinsic scatter in a satisfying way. In fact, we can go beyond
simple intrinsic scatter, and define a general unnormalized probability distribution
$\intdist(\vct{\para},\vct{\ortho})$ on $M$, adapted to
the relation $f$ through the coordinates $\surfchart$. This amounts to
setting up a $k$-dimensional probability distribution on $N_pS$ for each point
$p\in S$ and using the exponential mapping to create an $n$-dimensional 
distribution on $M$. However, as mentioned above, $\surfchart$ is not
injective when the geodesics normal to $S$ intersect each other. In these cases, all
the contributing points should be included, with corresponding
probabilities added, so that for a point $m\in M$ given in the original
coordinates, $\crdchart(m) = \vct{x}$, we have
\begin{equation}
    \intdist(\vct{x}) = \sum_{i\in I} \intdist(\vct{\para}_i(\vct{x}), \vct{\ortho}_i(\vct{x})),
\end{equation}
where
$I\subset\fZ_{+}$ indexes all the points on $S$ corresponding to
$m$. This definition is necessary for cases such as the wrapped normal
distribution on a circle $S^1$.

The distribution
$\intdist$ can represent e.g.\ a physical process producing objects with
correlated values of some physical parameters, but with intrinsic
scatter given by some probability distribution. \add{An} example would be the
process producing the correlation between the central black
hole mass and velocity dispersion in galaxies (see
Section~\ref{sc:msigma}).
The distributions of the possible physical measurements,
as observed in the chart $\crdchart$, are implicitly defined as the marginal
distributions of the distribution $\intdist$. These can be improper, as
can the distribution $\intdist$ itself.
That is, we may have
$\int_M \intdist\,\volf{\crdchart} = \infty$,
where
$\volf{\crdchart} = \sqrt{\abs{g}}\ud x^1\wedge\cdots\wedge\ud x^{n}$
is the volume form on $M$ in the chart $\crdchart$, and $g=\det(g_{ab})$.
\add{A concrete example of an improper $\intdist$ is given below.}

For applications, a useful distribution is the $k$-dimensional
normal distribution, independent of $\vct{\para}$, or the position on
$S$, defined by
\begin{equation}\label{eq:intndgaussian}
    \intdist(\vct{\ortho};\cvm) = 
    \frac{%
        \exp\left[-\frac{1}{2}\vct{\ortho}^{T}\cvm^{-1}\vct{\ortho}\right]
}{\sqrt{(2\pi)^k\det(\cvm)}},
\end{equation}
where $\cvm$ is the covariance matrix.
Of particular importance is the case where $k=1$, in which case we have
a one-dimensional normal distribution
depending only on $\ortho$, for which
\begin{equation}\label{eq:intgaussian}
    \intdist(\ortho; \sigma) =
    \frac{1}{\sqrt{2\pi\sigma^2}}\exp\left(-\frac{\ortho^2}{2\sigma^2}\right),
\end{equation}
where the parameter $\sigma$, standard deviation, is now an accurate
representation of what is typically called `intrinsic scatter' in
astrophysical literature.

\add{It should be noted that the
intrinsic scatter distributions \eqref{eq:intndgaussian} and \eqref{eq:intgaussian}
can yield a probability distribution that is not proper.  An example is
a line $y=\alpha + \beta x$ in flat two-dimensional space, for which, in
the coordinate chart, we have (for more details, see
Section~\ref{sc:lincase})
\begin{equation}
    \intdist(x,y) = 
    \frac{1}{\sqrt{2\pi\sigma^2}}\exp\left[-\frac{(y-\alpha-\beta
    x)^2}{2\sigma^2(1+\beta^2)}\right],
\end{equation}
for which $\int_{\fR^2} \intdist(x,y)\,\ud x\ud y$ diverges.
Similarly, in this case
the marginal distributions for $x$ and $y$, 
\begin{gather}
p_x(x)=\int_\fR \intdist(x,y)\,\ud y = \sqrt{1+\beta^2} \\
p_y(y)=\int_\fR \intdist(x,y)\,\ud x = \sqrt{1+\beta^2}/|\beta|,
\end{gather}
are similarly improper.
}
If necessary, \add{improper distributions $\intdist$ arising this way} can be
made proper e.g.\ with a suitable truncation in the coordinates $\vct{\para}$
followed by normalization. 

\begin{figure*}
    \begin{tabular}{ccc}
        \includegraphics[width=0.33\textwidth]{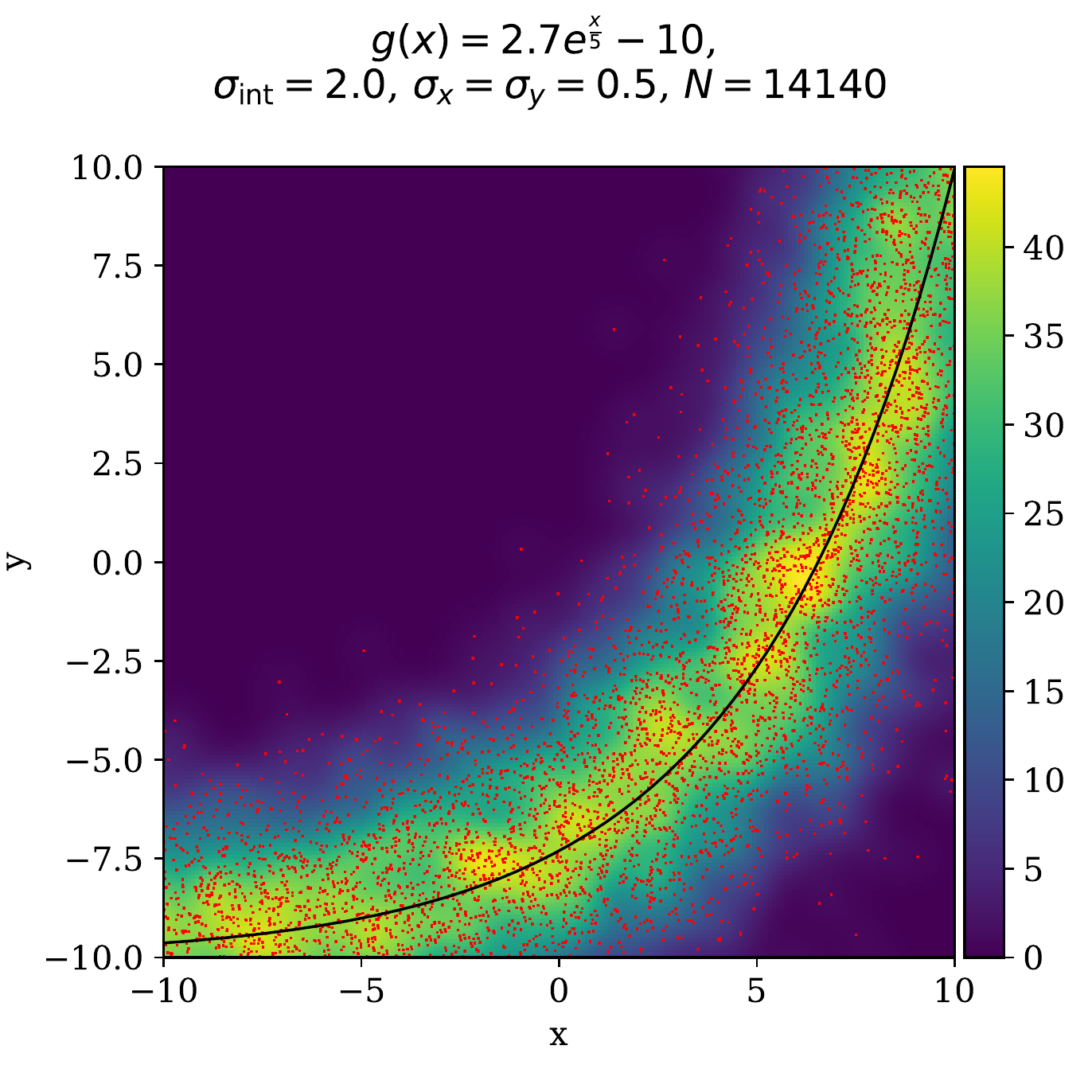} &
        \includegraphics[width=0.33\textwidth]{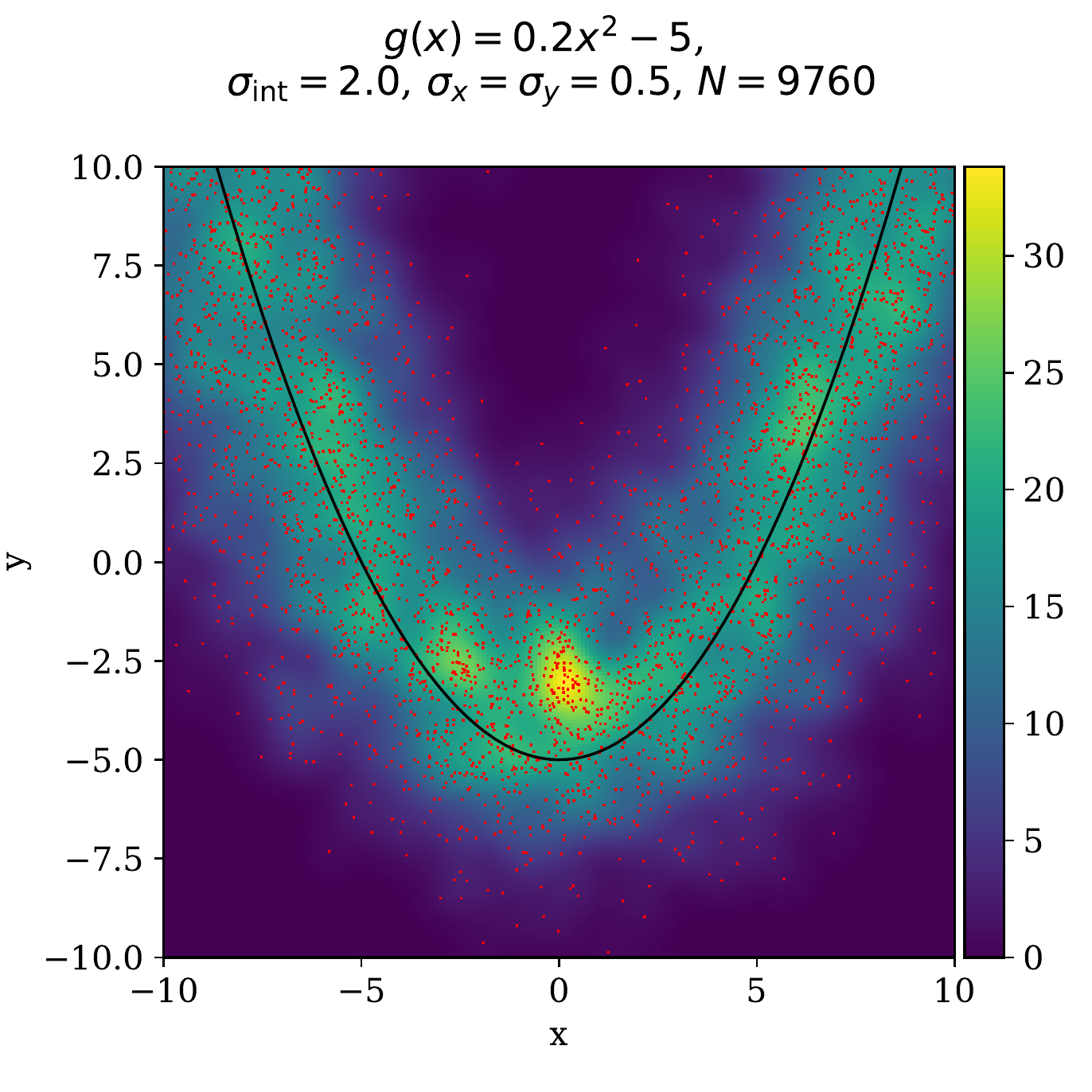} &
        \includegraphics[width=0.33\textwidth]{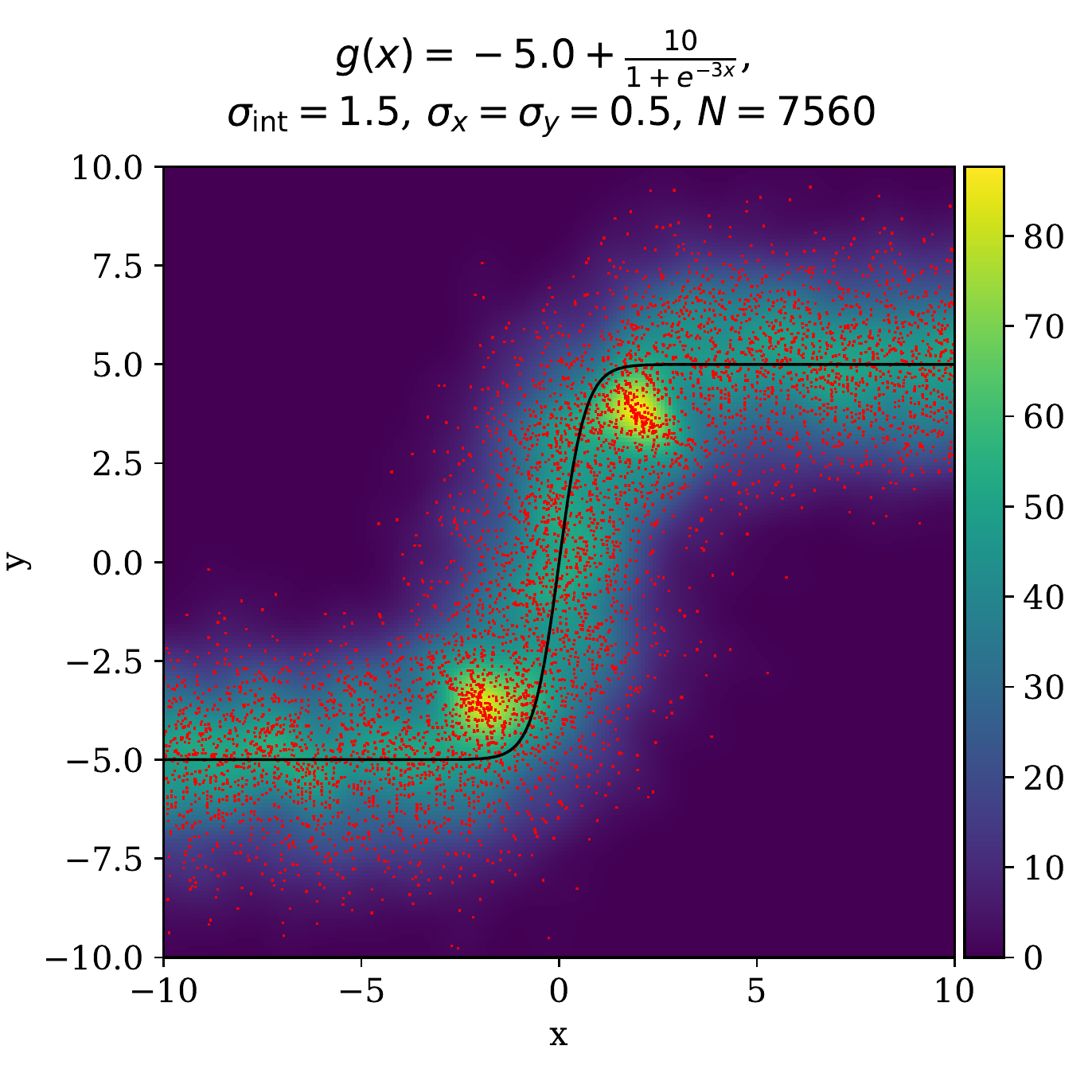}
    \end{tabular}
    \caption{Samples from a normally distributed $\intdist$,
    equation~\eqref{eq:intgaussian}, for non-linear one-dimensional relations
    $f(x) = y - g(x) = 0$ in $\fR^2$. The chosen relations are an exponential 
    (left), a parabola (middle) and a logistic curve (right).
The relations are shown in black, and the sampled points with red dots. 
The samples are given a two-dimensional zero-correlation Gaussian probability
distribution with standard deviations $\sigma_x$ and $\sigma_y$. The background
heatmaps show the summed sample probability density. The salient feature in the
images is the enhancement of the probability density towards the centre of
curvature, most prominently seen in the middle and right images.
}
    \label{fig:2d-dist-examples}
\end{figure*}

Figure~\ref{fig:2d-dist-examples} depicts 
samples of the distribution~\eqref{eq:intgaussian} for three non-linear
relations. From the figure it is easy to appreciate the fact that intrinsic
scatter can lead to distinct patterns of amplification and attenuation
in the probability density near the regions
where the underlying relation is strongly curved.
As such, a misleading result would be obtained from a fit to such data, if
the intrinsic scatter was not modelled properly, or at all.

\add{
Finally, we note that intrinsic scatter could also have been defined in
two other ways, either along coordinate directions or the scalar
geodesic distance to $S$. This seems to exhaust the possibilities, since
in general the only directions we have available are the
coordinate directions, or alternatively directions along the relation
and orthogonal to the direction. In addition, only geodesic distance
is a sensible natural choice in a Riemannian context. 
It should be emphasized that the definitions of intrinsic scatter using
the intrinsic coordinates $\surfchart$ or the coordinate chart
$\crdchart$ are in principle completely equivalent, and merely
formulated in different coordinates. Both are more general than a
definition based only on geodesic distance from the relation $S$.
}

\add{
A definition based on coordinate directions is often used at least in
astronomical literature, where in the case of a line in two dimensions
$(x,y)$, the intrinsic scatter is typically taken to lie in the
direction of the $y$-coordinate \citep[e.g.][]{press1992,tremaine2002,gultekin2009}. 
This gives a relation that is `puffed
up' in the $y$-direction.
This approach has the obvious problem of not being
invariant with respect to a change of coordinates.
However, it may be reasonable in some cases, where we
are certain that whatever processes are producing the intrinsic scatter
only operates strictly along a particular measured observable.
Nevertheless, considering
the often very correlated nature of astronomical observables,
establishing this fact seems to be a difficult proposition. Consider for
example the \msigma relation between the mass of a supermassive black
hole and the central velocity dispersion of its host galaxy, where the
intrinsic scatter is typically taken to lie entirely along the
$M_{\text{BH}}$ axis.
As galaxies
together with their SMBHs evolve, they move on the \msigma plane, but
with different galaxies starting at (nearly) the same point, not
necessarily ending up near each other later due to differences in
composition, environment, merger history and so on. This produces
intrinsic scatter. However, it seems very difficult to justify that
these processes should operate only along the black hole mass axis.
Rather, it seems that they might operate in any which direction
indeterminately, but of these, the only one evident to us is \emph{the
direction away from the underlying relation}. Indeed, if the processes
producing intrinsic scatter would only operate along the relation,
the resulting data would have no scatter around the relation at all.
Finally, comparing the magnitude of intrinsic scatter between datasets
becomes difficult, since the value will end up depending on the
orientation of the line (see Section~\ref{sc:comparison}).
In this sense, it seems reasonable to formulate intrinsic scatter
problems in terms of the intrinsic coordinates $\surfchart$,
separately considering the directions orthogonal and parallel to the
proposed relation.
}

As the last option, the intrinsic scatter could also have been defined as
a probability density in the original measurement coordinate system
$\crdchart$, depending only on the scalar geodesic distance to $S$.
This approach
would give relations that are `puffed up' symmetrically in all
directions, with density falling off as a given function
of the geodesic distance.
Probability densities using this approach would not present the
enhancements shown in Figure~\ref{fig:2d-dist-examples}. \add{However, this
approach is strictly less general than the one outlined above, and}
does not allow for e.g.\ wrapped distributions on spheres,
toroids and so on, and cannot cope with directionally varying scatter
for submanifolds $S$ with codimension higher than one.

\subsection{Fitting a relation with intrinsic scatter}\label{sc:intscat}

In a typical case, we have obtained $n_d$ measurements $\data =
\{(m_1,h_1),\ldots,(m_{n_d},h_{n_d})\}$, where $m_i\in M$ are the
measured values and $h_i:M\fromto\fR$ are the probability distributions
$p(\eta_i|m_i)$ for the true value $\eta_i$, given the measurement
$m_i$, typically representing the estimated measurement errors.
The $h_i$ are otherwise unspecified, and
arbitrary distributions of measurement errors, including upper or lower
limits (left or right censoring, respectively) can be naturally
accommodated.
For optimal results, the distributions $h_i$ should be Bayesian
posterior probability densities obtained in the measurement process,
fully representing the available information. However, in some cases
$h_i$ are completely unknown, in which case they must be estimated,
typically as Gaussians with unknown covariance matrices, which then must
taken as additional nuisance parameters, in addition to the true values
$\eta$.
In addition to the measurements, we have
a postulated relation $f$, parameterized by $n_p$ parameters
$\relpars\in\fR^{n_p}$, from which we derive coordinates
$\surfchart(m;\relpars)$, also parameterized by $\relpars$.
We believe the data to fulfil the relation $f$, up to some
intrinsic scatter $\intdist$, parameterized by $n_s$ parameters
$\intpars\in\fR^{n_s}$. We would now like to find the most
probable values of $\relpars$ and $\intpars$, constrained by
the data $\data$. 

The Bayesian way to proceed is to note that for a single datum
$(m,h)$, the conditional probability of the measurement factors as
$p(m,h|\relpars,\intpars) = \int
p(m,h|\eta)\intdist(\eta|\relpars,\intpars)\,\ud\eta$,
where $\eta$ is the true value, drawn from the intrinsic distribution, $\intdist$. 
As such, the true value $\eta$ is taken as a nuisance
parameter, and integrated out.
In addition, it may be that the measurement process is not fully
sensitive across the range of possible values, leading to truncation, or
data points that are not seen in the sample. If the probability that a
given value $\eta$ leads to a detection is given by
$\detdist(\eta)$, we have for the measurement probability that
$p(m,h|\relpars,\intpars) = \int p(m,h|\eta)\detdist(\eta)\intdist(\eta|\relpars,\intpars)\,\ud\eta$.
See Appendix~\ref{sc:appendix-a} for full derivation.

Now, taken as a function of the parameters, the probability
$p(\data|\relpars,\intpars)$ defines the likelihood $\mathcal{L}$ of $\relpars$
and $\intpars$. Assuming that the data $\data$ are independent, we get
\begin{equation}\label{eq:likelihood}
    \begin{split}
        &\mathcal{L}(\relpars,\intpars|\data) = \\
        &\,
        \prod_{i=1}^{n_d} \int_M\!\!\!\!
        h_i(\vct{x}(\vct{\para},\ortho;\relpars))
        \detdist(\vct{x}(\vct{\para},\ortho;\relpars))
        \intdist(\vct{\para},\ortho;\intpars)
        \,\volf{\surfchart} = \\
        &\,
        \prod_{i=1}^{n_d} \int_M\!\!\!\!
        h_i(\vct{x})
        \detdist(\vct{x})
        \sum_{j\in I}\intdist(\vct{\para}_j(\vct{x};\relpars),\vct{\ortho}_j(\vct{x};\relpars);\intpars)
        \,\volf{\crdchart},
    \end{split}
\end{equation}
where the integration can be done either in the coordinate chart
$\crdchart$ or the intrinsic chart $\surfchart$. 
The integration in $\surfchart$ automatically takes
care of points $m\in M$ that have multiple representations
$(\vct{\para}_j,\vct{\ortho}_j)$, but if the integration is done over
$\crdchart$, these have to be explicitly summed over.

If $\relpars$ and $\intpars$ have a joint prior
distribution $\pi_{\relpars,\intpars}$ (possibly uninformative),
we can define the joint posterior
probability distribution of $\relpars$ and $\intpars$,
\begin{equation}\label{eq:postdist}
    \postdist(\relpars,\intpars) =
    \frac{
        \mathcal{L}(\relpars,\intpars|\data)\pi_{\relpars,\intpars}(\relpars,\intpars)
    }{%
        \iint
        \mathcal{L}(\relpars',\intpars'|\data)\pi_{\relpars,\intpars}(\relpars',\intpars')
        \,\ud\relpars'\ud\intpars'
        }.
\end{equation}
The posterior distribution, equation \eqref{eq:postdist}, contains all
the information of how the data constrains the parameters $\relpars$ of the
relation itself, as well as the
parameters $\intpars$ of the intrinsic scatter model, given our prior
information $\pi_{\relpars,\intpars}$. 
In some special cases, the
posterior distribution can be computed analytically, but in general it
is necessary to employ numerical techniques, \add{such as a suitable
Markov Chain Monte Carlo (MCMC) method}
\st{such as Gibbs sampling}
\citep[see e.g.][]{gelman2013}. 
\add{Finally, point estimates such as the Maximum (Log-)Likelihood
Estimate (MLE) and Maximum A Posteriori Estimate (MAP) can be obtained
using the likelihood via standard procedures.}
Full derivation of the
likelihood~\eqref{eq:likelihood} and the posterior
distribution~\eqref{eq:postdist} following the style of
\citet{jaynes2003} is found in Appendix~\ref{sc:appendix-a}.


\subsection{Some considerations}\label{sc:some_considerations}

For the formulation presented above to make sense, a reasonable geometry
must exist between the measured quantities. This is not a problem e.g.\ 
for measurements of a 3D angle, where we have the natural geometry of the
sphere $S^2$.
\add{In other cases, such as for} joint measurements of
galaxy luminosity and velocity dispersion it is reasonable to question
whether a distance measured `across' the measured quantities makes any
sense. This observation has been discussed numerous times since Pearson
in 1901 originally presented a least squares method using orthogonal distances
to the regression line (or plane), typically called Orthogonal Regression (OR)
\citep{pearson1901}. See for example \citet{isobe1990} for a discussion.

The Riemannian approach presented in this paper makes it possible to
specify the units and any possible geometry between the measured
quantities beforehand. After the geometry is set, the metric character
of the approach \add{ensures} that results are invariant with respect to
all coordinate transformations, which in this case would typically
represent scaling of the measured quantities (i.e.\ a change of units).
Another natural possibility is to introduce the measurements in an
entirely scale-free manner by using logarithms.\footnote{
Linear measurements $x^i$ with a metric $g=\diag(1,\ldots,1)$ 
are equivalent to logarithmic measurements $\log_{10} x^i$ and a metric
with an exponential dependence on the measured values, 
$g=\log(10) \diag(10^{x^1},\ldots,10^{x^n})$.}

A practical problem is the fact that the coordinate transformation between
$\crdchart$ and $\surfchart$ may prove difficult or impossible to derive
analytically. Finding the transformation and computing the posterior
probabilities numerically necessitates computing numerous geodesic distances
in a curved space. In general this is highly computationally demanding, although
efficient algorithms exist for some special cases \citep{crane2013}.

\add{Finally, we note that there are to be some consequences from and
natural restrictions to the choice of parameterization and the
prior distributions of the parameters. Firstly, it is clear that 
an arbitrary transformation of parameters
$\vct{\theta}=g(\vct{\theta}')$ will result in the change of
the prior probability measure so that 
$\pi(\vct{\theta})\ud\vct{\theta}=\pi(g(\vct{\theta}'))\det(J_g)\ud\vct{\theta}'$,
where $J_g$ is the Jacobian of the transformation. This can be
understood also through the fact that a probability distribution $\pi$ 
defined in the parameter space also yields through the relation $f$ and
$\intdist$ a probability distribution on $M$. This latter distribution,
once defined, should naturally be invariant under a change of
parameterization, which is guaranteed by the transformation formula
above.
}

\add{
Secondly, prior distributions are usually considered to be somewhat
arbitrary and application-specific, often chosen for convenience to
yield analytic results for the posterior probability. However, as noted
above, this freedom can be restricted by what we know of how the data is
distributed on $M$ combined with the choice of a relation $f$ and
$\intdist$. An important special case is if we have no prior knowledge
of how the data is distributed, but only the possible values that it may
take, through the choice of $M$. In this case the maximum entropy
distribution on $M$ is uniform. This distribution places restrictions on
the choice of $f(\vct{x};\vct{\theta})$, and after $f$ is chosen, forces
a particular choice of $\pi(\vct{\theta})$, which can then in this
context be called \emph{the} uninformative prior.
The general argument is outside the scope of this paper, but see e.g.\ 
\citet{george1993}, \citet{broemeling2003} and \citet{fraser2010} for
historical and modern discussions on
vague and uninformative priors from different points of view.
However, a special case of an uninformative prior defined in the way
described above is considered in detail in Section~\ref{sc:lincase} and
later used in Sections \ref{sc:comparison} and \ref{sc:msigma}.
}

\section{Worked out examples}\label{sc:specialcases}

\add{The following sections contain analytical results for 
particular choices of the measurement manifold $M$ and the relation $f$.
In each case, we show how the framework presented above is used to
obtain the likelihood, which can then be used for parameter estimation.
The outline of the process, or a `practical how-to', is as follows:
\begin{enumerate}
    \item Determine the manifold $M$, along with the metric $g$. These
        are typically fixed by the nature of the measured
        quantities.
        \label{todo:pt1}
    \item Fix the functional form $f(x;\relpars)$ of the relation to be
        fit, along with the parameters $\relpars$ of the relation.
        \label{todo:pt2}
    \item Define the intrinsic coordinates $(\para,\ortho)$.
        A well-motivated choice for the coordinates $\para$ along the
        relation is using geodesic distances from a given fixed origin.
        The coordinates $\ortho$ are then defined up to a rotation of
        the $\ortho$-coordinate axes.
        \label{todo:pt3}
    \item Determine the coordinate transformation
        \mbox{$\crdchart\circ\surfchart^{-1}$} from $(\para,\ortho)$ to $x$ or
        the inverse transformation
        \mbox{$\surfchart\circ\crdchart^{-1}$}.
        This may only be possible analytically in one direction
        only, or may not be possible analytically at all.
        \label{todo:pt4}
    \item Choose a suitable intrinsic scatter model
        $\intdist(\ortho,\para)$ as a
        function of the intrinsic coordinates.
        \label{todo:pt5}
    \item Evaluate the likelihood integral,
        equation~\eqref{eq:likelihood}. For analytical results, this
        typically needs to be done only for a single arbitrary
        datapoint, if all the datapoints have a similar error
        distribution model.
        \label{todo:pt6}
    \item Compute the likelihood over all data as a product of the
        likelihood of the individual likelihoods for each data point.
        Equivalently, choose a suitable prior and
        compute the posterior distribution,
        equation~\eqref{eq:postdist}, over all the data.
        \label{todo:pt7}
\end{enumerate}
We will refer to these steps in the following derivations so that the
process can be more easily followed.
}

\subsection{Linear $n-1$-dimensional case}
\label{sc:nd-case}

A highly useful special case is the case of normally distributed
intrinsic scatter, equation \eqref{eq:intgaussian}, in a linear relation
$f:\fR^n\fromto\fR$ between $n$ observables, defining an
$n-1$-dimensional (codimension $1$) hyperplane $S$, with measurements
drawn from an Euclidean space $M=\fR^n$. \add{For now, we assume that the
metric $g$ is given by the $n\times n$ identity matrix $\mat{I}$.
These definitions complete step \ref{todo:pt1} from above.}
This special case represents well the
typical astrophysical problem of modelling a linear relation between
multiple physical variables that have \st{a} no special geometry, such as
mass, luminosity or velocity dispersion. 

\add{To address step \ref{todo:pt2}, we fix the parametrization.}
All \add{the} planes $S$ can be parameterized with
$\vct{\relpars}=\vct{p}\in\fR^n$ so that 
\begin{equation}
    f(\vct{x};\vct{\relpars}) = (\vct{x}-\vct{p})^T \vct{p}
    = \vct{x}^T \vct{p} - \norm{\vct{p}}^2 = 0,
\end{equation}
where $\vct{x}\in \fR^n$.
With this parameterization, the vector $\vct{p}$ corresponds to the
point of $S$ where it is closest to the origin, and as such $\vct{p}$
is also normal to the plane $S$. A different parameterization, more often used in
the astrophysical literature, emphasizes one particular coordinate dimension,
written as
\begin{equation}
    x_1 = a_1 + \sum_{j=2}^{n}a_i x_i.
\end{equation}
Converting between these two parameterizations is accomplished through
\begin{alignat}{3}
    a_1 &= \frac{\sum_{i=1}^{n} p_{i}^2}{p_{1}}, &\quad a_i &= -\frac{p_{i}}{p_{1}}, &\quad& i=2,\ldots,n \\
    p_{1} &= \frac{a_1}{1+\sum_{i=2}^{n} a_i^2}, &\quad p_{i} &= -a_i p_0,           &\quad& i=2,\ldots,n.
\end{alignat}

We further assume that for each measurement $(\vct{m},h)$, where now
$\vct{m}\in\fR^n$, the measurement error can be modelled with an
$n$-dimensional normal distribution
\begin{equation}\label{eq:error-nd-gaussian}
    h(\vct{x};\vct{m},\cvm) = \frac{%
    \exp\left[-\frac{1}{2}(\vct{x}-\vct{m})^{T}\cvm^{-1}(\vct{x}-\vct{m})\right]
}{\sqrt{(2\pi)^n\det(\cvm)}},
\end{equation}
where $\cvm$ is the measurement error covariance matrix. This assumption
is necessary to obtain the analytic results below, but it is also
well suited to published astrophysical data, for which the
complete posterior distributions for each datum are typically not
available. \add{Upper limits can be incorporated by substituting
one or more degrees of freedom $x_i$ in equation~\eqref{eq:error-nd-gaussian}
with suitable one-dimensional distributions representing the limit
distributions.}

\add{To complete steps \ref{todo:pt3} and \ref{todo:pt4}, we need to
specify the intrinsic coordinates and the coordinate transformations.}
One method to construct the coordinates $\surfchart$ is to transform
the natural coordinate frame of the measurement space $M$ with a
translation by $\vct{p}$ and a 
rotation $\mat{R}\in SO(n)$, which \st{then} takes the $n$'th basis vector $\hat{\vct{e}}_n$
to the unit vector $\hat{\vct{p}}$ (hereafter, we use a hat to signify a unit vector). 
The new components of
a vector $\vct{x}$ are then 
$\surfchart(\vct{x}) = \mat{R}^T(\vct{x}-\vct{p}) = (\vct{\para},\ortho)\in\fR^{n-1}\times\fR$.
Defining $\vct{w} = (p_{0},\ldots,p_{n-1},0)$, $\cos\theta =
\hat{\vct{p}}^T\hat{\vct{e}}_n$ and $\sin\theta =
\hat{\vct{p}}^T\hat{\vct{w}}$, we can compute $\mat{R}$ with
\begin{equation}
    \mat{R}(\vct{p}) = I - \hat{\vct{w}} \hat{\vct{w}}^T - \hat{\vct{e}}_n \hat{\vct{e}}_n^T
    + \left(\hat{\vct{w}}\;  \hat{\vct{e}}_n\right) 
    \mat{R}_2(\theta)
        \left(\hat{\vct{w}}\;  \hat{\vct{e}}_n\right)^T,
\end{equation}
where
\begin{equation}
    \mat{R}_2(\theta) = 
    \begin{pmatrix}
        \cos\theta & \sin\theta \\
        -\sin\theta & \cos\theta
    \end{pmatrix}
\end{equation}
is the usual $2$-dimensional rotation matrix. \add{The origin of the
$\vct{\para}$ coordinates is then at $\vct{p}$ (in chart $\crdchart$), 
with the orientation of
the coordinate axes $\hat{\vct{\para}}_i$ given by $\hat{\vct{\para}}_i
= \mat{R}\hat{\vct{e}_i}$, for $i=1,\ldots,n-1$. The orthogonal
intrinsic coordinate axis is $\hat{\vct{\ortho}}_1 =
\mat{R}\hat{\vct{e}}_n$.
}

\add{Step \ref{todo:pt5} requires specification of the intrinsic
scatter model $\intdist$. We assume the normal intrinsic distribution,
equation~\eqref{eq:intgaussian}. With this definition, combined with the
assumed measurement error distribution, and
assuming no censoring, that is no upper or lower limits, we can complete
step \ref{todo:pt6}. We evaluate
the equation~\eqref{eq:likelihood} for the likelihood of a single
measurement datapoint.} The result is
\begin{equation}\label{eq:ndimllh}
    \mathcal{L}(\vct{p},\sigma|\vct{m},\cvm) = 
    \frac{1}{\sqrt{2\pi(\tilde{\Sigma}_{nn} + \sigma^2)}}
    \exp\left[\frac{-\tilde{m}_n^2}{2(\tilde{\Sigma}_{nn}+\sigma^2)}\right],
\end{equation}
where 
\begin{align}
    \tilde{\vct{m}} &= \mat{R}^T(\vct{m}-\vct{p}) \\
    \tilde{\cvm} &= \mat{R}^T\cvm \mat{R}.
\end{align}
We can also write $\tilde{m}_n = \hat{\vct{p}}^T\vct{m} - \norm{\vct{p}}$
and $\tilde{\Sigma}_{nn} = \hat{\vct{p}}^T\cvm\hat{\vct{p}}$,
which transforms equation~\eqref{eq:ndimllh} to the form used in
\citet{robotham2015}.

\add{We can also address the case where the measurements are given in
coordinates where the metric is represented by a \emph{constant} matrix
$\mat{G}\neq \mat{I}$.
This can be the case when the metric $\mat{G}$ encodes the choice of
units for the variables, or when it has off-diagonal terms. 
A situation that would lead to a non-diagonal metric is, for example, 
a case where we are interested in two observables $A_1$ and $A_2$, 
but can only measure $B_1=A_1$ and $B_2=A_1+A_2$. The observables may be
e.g.\ inflows of current or liquid from two independent sources, but
we can only measure one source directly, and the sum of the
flows somewhere downstream. The flows $A_i$ can by assumption
have any values, so the metric in coordinates $A_i$ should be a diagonal
product metric, whereas in coordinates $B_i$ off-diagonal terms appear.
For the case of a non-identity constant metric $G$, the equation}
\eqref{eq:ndimllh} is still valid, but we have
\begin{align}
    \tilde{\vct{m}} &= \mat{R}^T \mat{W} \mat{P}^T(\vct{m}-\vct{p}) \\
    \tilde{\Sigma} &=  \mat{R}^T \mat{W} \mat{P}^T\Sigma \mat{P} \mat{W}
    \mat{R},
\end{align}
where $\mat{W}$ is a diagonal matrix of the square roots of the eigenvalues of
$\mat{G}$, and $\mat{P}\in O(n)$, so that
$\mat{W}^{-1}\mat{P}^T \mat{G} \mat{P} \mat{W}^{-1} = \mat{I}$. This is
possible, since the metric is assumed to be Riemannian, in which case
all the eigenvalues of $\mat{G}$ must be positive. 
\add{Note that if the metric is not constant, the shortcut presented
above does \emph{not} work, and all the steps
\ref{todo:pt1}-\ref{todo:pt7} have to be followed using the metric
explicitly.}


\subsection{Linear $n-k$-dimensional case}\label{sc:n-k-case}

The approach in the previous section can be readily generalized to the
case of $k>1$ simultaneous linear relations, defining an
$n-k$-dimensional (codimension $k$) affine subspace $S_{n-k}$.
\add{In this case, it is easiest to define the relation through defining
the intrinsic coordinates first. As such, we go through steps
\ref{todo:pt3} and \ref{todo:pt4} first.}
This can be accomplished by
starting with the $n-1$-dimensional subspace $S_{n-1}$ defined by the
first relation, through the $n$ parameters $\vct{p}_n\in\fR^n$, as above. 
Now, the subspace $S_{n-k}$ must lie in the intersection of
$S_{n-1}$ and an $n-2$-dimensional subspace $S_{n-2}$, which itself must
also lie entirely in $S_{n-1}$. We may parameterize $S_{n-2}$ \emph{within
$S_{n-1}$} with the $n-1$ parameters
$\vct{p}_{n-1}=(p_{n-1,1},\ldots,p_{n-1,n-1},0)\in\fR^n$. This process
is then continued until we have given $p_{n-k+1}$, parameterizing
$S_{n-k}$ in $S_{n-k+1}$. The number of parameters is $(2nk-k^2+k)/2$ in total.
Each parameter vector $\vct{p}_i$ yields a
rotation matrix $\mat{R}_{i}$ as in the section above, which
together with $\vct{p}_i$ defines the coordinate transformation from the
intrinsic coordinates $\surfchart$ on $S_{i}$ to the intrinsic
coordinates on $S_{i-1}$, with the understanding that $S_{n}=M$.
Following this process through, we find that the coordinate
transformation from $\crdchart$ to the intrinsic coordinates on
$S_{n-k}$ is given by
\begin{equation}
    \begin{split}
        (\surfchart\circ\crdchart^{-1})(\vct{x}) &= 
    \mat{R}_{n-k+1}^T\cdots\mat{R}_{n}^T(\vct{x} - \vct{p}_{n}) \\
        &\quad 
        -\mat{R}_{n-k+1}^T\cdots\mat{R}_{n-1}^T\vct{p}_{n-1}
    - \cdots \\
        &\quad
    -\mat{R}_{n-k+1}^T\vct{p}_{n-k+1} \\
        &= \mat{R}^T(\vct{x}-\vct{p}) \\
        &= (\vct{\para},\vct{\ortho}) \in \fR^{n-k}\times\fR^{k},
    \end{split}
\end{equation}
where
\begin{equation}
    \mat{R}^T = \mat{R}_{n-k+1}^T\cdots\mat{R}_{n}^T
\end{equation}
and
\begin{equation}
    \vct{p} = \vct{p}_{n} + \mat{R}_{n}\vct{p}_{n-1} + \cdots + \mat{R}_{n}\cdots\mat{R}_{n-k+2}\vct{p}_{n-k+1}.
\end{equation}
\add{Again, the origin of the intrinsic coordinate axes is at $\vct{p}$.
The orientation is similarily given by $\hat{\vct{\para}}_i = \mat{R}\hat{\vct{e}}_i$
for $i=1,\ldots,n-k$ and $\hat{\vct{\ortho}}_i =
\mat{R}\hat{\vct{e}}_{n-k+i}$ for $i=1,\ldots,k$. This process defines the
relation $f$ in a roundabout way as the equation $\vct{\ortho}=\vct{0}$, which
in the chart $\crdchart$ is then given by the system of $k$ equations
\begin{equation}
    f(\vct{x};\vct{\relpars}) =
    [\mat{R}^T(\vct{x}-\vct{p})]_{n-k+1,\ldots,n} = \vct{0},
\end{equation}
where $\relpars=(\vct{p}_{n},\ldots,\vct{p}_{n-k+1})$.
This completes step \ref{todo:pt2}.
}

We again assume that measurement errors are normally distributed,
given by equation~\eqref{eq:error-nd-gaussian}.
\add{For step \ref{todo:pt5} we now assume that the intrinsic scatter is normally distributed in
$\vct{\ortho}$ with a covariance matrix $\intcvm$, as in
equation~\eqref{eq:intndgaussian}.
With this definition, we can complete step \ref{todo:pt6} and compute the likelihood given
by a single measurement $\vct{m}$.} The result is
\begin{equation}\label{eq:nk-likelihood}
    \mathcal{L}(\vct{\relpars},\intcvm|\vct{m},\cvm)
    = \frac{%
        \exp\left[-\frac{1}{2}
        \tilde{\vct{m}}_\ortho^T(\tilde{\cvm}_{\ortho\ortho}+\intcvm)^{-1}\tilde{\vct{m}}_\ortho
        \right]
        }{%
            \sqrt{(2\pi)^k \det\left(\tilde{\cvm}_{\ortho\ortho} + \intcvm\right)}
        },
\end{equation}
where
\begin{align}
    \tilde{\vct{m}} &= \mat{R}^T(\vct{x}-\vct{p}) = (\tilde{\vct{m}}_\para,\tilde{\vct{m}}_\ortho) \\
    \tilde{\cvm} &= \mat{R}^T\cvm\mat{R} = 
    \begin{pmatrix}
        \tilde{\cvm}_{\para\para} & \tilde{\cvm}_{\para\ortho} \\
        \tilde{\cvm}_{\ortho\para} & \tilde{\cvm}_{\ortho\ortho}
    \end{pmatrix}.
\end{align}
A possible constant non-Euclidean metric can be accommodated as in the previous section. 


The likelihood, equation~\eqref{eq:nk-likelihood} is readily generalized for
intrinsic distributions other than the multivariate normal
distribution through mixture models \citep[see e.g.\ the approach
in][]{kelly2007}.
In addition, the result in this
section is useful for hypothesis testing in the sense of finding the
most likely value of $k$ for a given $n$-dimensional dataset.

\subsection{The line in two dimensions}\label{sc:lincase}

It is useful to work out the two-dimensional special case of a line with intrinsic
scatter in detail,
considering the amount of
literature focusing on symmetric fitting of linear relations with and
without intrinsic scatter
\citep[e.g.][and many
others]{pearson1901,boggs1987,isobe1990,feigelson1992,robotham2015}.
\add{As such, step \ref{todo:pt1} consists of setting $M=\fR^2$ with an
Euclidean metric. In step \ref{todo:pt2},}
for ease of comparison with existing methods, instead of the parameterization
used above, we define the relation $f$ and the line $S$ with
\begin{equation}
    f(x,y;\vct{\relpars}) = y - \beta x - \alpha = 0,
\end{equation}
\add{where now $\vct{\relpars} = (\alpha,\beta)\in\fR^2$.}
The measurement error in the case of no censoring can now be represented
with a two-dimensional normal distribution, given by
\begin{equation}\label{eq:2dgaussian}
    \begin{split}
        &h(x,y; x_0, y_0, \sigma_x, \sigma_y, \rho) = \\
        &\quad
        \frac{%
        \exp\left\{
            -\frac{1}{2(1-\rho^2)}
            \left[
                \frac{(x-x_0)^2}{\sigma_x^2}
                + \frac{(y-y_0)^2}{\sigma_y^2}
                - \frac{2\rho(x-x_0)(y-y_0)}{\sigma_x\sigma_y}
            \right]
        \right\}
        }
        {2\pi\sigma_x\sigma_y\sqrt{1-\rho^2}},
    \end{split}
\end{equation}
where $x_0$ and $y_0$ specify the measured values, $\sigma_x$,
$\sigma_y$ represent the measurement uncertainties in the $x$ and $y$
directions, and $\rho\in[-1,1]$ specifies the correlation between the
measurement errors. Upper and lower limits can be introduced as in
Section~\ref{sc:nd-case}. While the parameterization through $\alpha$ and $\beta$
is convenient and intuitive, it is not manifestly symmetric with respect to the
coordinates. This will be investigated further below.

Keeping the assumption that the internal
scatter is normally distributed, via equation~\eqref{eq:intgaussian}, 
\add{we can use the results for steps \ref{todo:pt3} to \ref{todo:pt6}
from Section~\ref{sc:nd-case}.}
The likelihood of a single measurement in the case of no censoring is
then obtained from equation~\eqref{eq:ndimllh}, yielding
\begin{equation}\label{eq:llh-single}
    \mathcal{L}(\alpha,\beta,\sigma|x_0,y_0,\sigma_x,\sigma_y,\rho) =
    \frac{1}{\sqrt{2\pi\tilde{\sigma}^2}} \exp\left(-\frac{\ortho^2}{2\tilde{\sigma}^2}\right),
\end{equation}
where now
\begin{align}
    \ortho^2 &= \frac{(y_0 - \alpha - \beta x_0)^2}{1+\beta^2}\label{eq:ab-delta} \\
    \tilde{\sigma}^2 &= 
    \frac{\beta^2\sigma_x^2 + \sigma_y^2 - 2\beta\rho\sigma_x\sigma_y}{1+\beta^2} + \sigma^2, \label{eq:ab-sigma}
\end{align}
so that $\ortho^2$ is the squared orthogonal distance from the relation, and
$\tilde{\sigma}$ is an extended uncertainty incorporating both intrinsic scatter
and the measurement errors. \add{Analytic results for particular
censored error distributions can also be found. See
Appendix~\ref{sc:appendix-b}.}

We note that equation~\eqref{eq:ab-sigma} has the correct asymptotic
behaviour with respect to $\beta$, in the sense that when the regression line
tends towards the vertical, or $\beta\fromto\infty$, we have
$\tilde{\sigma}\fromto \sigma_x^2+\sigma^2$, agreeing with intuitive result that
all of the uncertainty should in this case be a combination of the horizontal
(along $x$-axis) and intrinsic scatter.
Likewise, when the regression line tends towards the horizontal, or $\beta
\fromto 0$, we have $\tilde{\sigma}\fromto \sigma_y^2+\sigma^2$, similarly
agreeing with geometric intuition.

\add{
We now apply the discussion in Section~\ref{sc:some_considerations}
and derive a strictly unique uninformative prior probability density for the
parameters $\relpars=(\alpha, \beta)$.
This is appropriate for the case where no data has yet been collected,
and we have no information on where on the plane the relation is and in
what orientation. Thus, we need to find a prior density for $\relpars$
which yields a uniform distribution on $M$ when integrated over. That
is, the lines must cover the plane evenly. A direct approach seems to
necessitate set-based analysis, but in this special case we can apply
existing results from the literature, derived via other means.
}

\add{
For example, demanding invariance under Euclidean coordinate transformations
(simultaneous rotation and translation) yields an invariant prior
probability measure $\pi(\alpha,\theta)\ud\alpha\ud\theta =
\ud\alpha\ud\sin\theta$, where $\theta=\arctan\beta$ is the angle between the line
and the $x$-axis. This is equivalent to
$\pi(\alpha,\beta)=(1+\beta^2)^{-3/2}$.
The same result can be obtained by demanding that the prior density be invariant under a switch
of the coordinates, that is under
$x\leftrightarrow
y$, $\sigma_x\leftrightarrow\sigma_y$,
$\beta\leftrightarrow 1/\beta$ and $\alpha\leftrightarrow -\alpha/\beta$.
This result was apparently originally derived by E.T.~Jaynes in 1976
(reprinted in \citealt{jaynes1983}).
}

\add{
Finally, for $\sigma$ there is considerably more leeway in the
literature with regards to the choice of an uninformative prior.
A typical choice is the Jeffreys prior, $\pi(\sigma)\propto 1/\sigma$. 
The complete
uninformative prior for this special case is then
\begin{equation}\label{eq:2d-prior}
    \pi_{\intpars,\relpars}(\alpha,\beta,\sigma) = \frac{1}{\sigma (1+\beta^2)^{3/2}}.
\end{equation}
}


Equation \eqref{eq:2d-prior} and
\eqref{eq:llh-single} can now be combined to yield the unnormalized posterior
distribution 
\begin{equation}\label{eq:2d-post}
    p(\alpha,\beta,\sigma|\data) =
    \frac{1}{\sigma(1+\beta^2)^{3/2}}
    \prod_{i=1}^{n_d}\mathcal{L}(\alpha,\beta,\sigma|x_i,y_i,\sigma_{x,i},\sigma_{y,i},\rho_i).
\end{equation}
This gives a Bayesian solution to the
problem of symmetric fitting of a linear relation with intrinsic scatter
to two-dimensional data with heteroscedastic errors in both measured
variables, including possible upper or lower limits and truncation. 
This result thus
extends the earlier Bayesian results in \citet{zellner1971},
\citet{gull1989}, \citet[][unpublished]{jaynes1991} and \citet{kelly2007}.
It should be
noted that while the likelihood, equation \eqref{eq:llh-single} is
invariant under the switch of coordinates,
the posterior distribution in itself is not, \add{if the Jacobian of the
transformation of parameters is not included, as discussed in
Section~\ref{sc:some_considerations}.}
In practice this means that in the usual case of $f=y-\alpha-\beta
x$ we wish to find the maximum of
\begin{equation}
    p(\alpha,\beta,\sigma|x,y,\sigma_x,\sigma_y) = p(\alpha,\beta,\sigma)\mathcal{L}(\alpha,\beta,\sigma|x,y,\sigma_x,\sigma_y).
\end{equation}
In the inverse case, where $f = x - \alpha' - \beta' y$, and
$\alpha'=-\alpha/\beta$, $\beta'=1/\beta$, we should maximize
\begin{equation}
\begin{split}
    &\abs{\beta'^3}p(\alpha',\beta',\sigma)\mathcal{L}(\alpha',\beta',\sigma|y,x,\sigma_y,\sigma_x) = \\
    &\quad
    p(\alpha,\beta,\sigma)\mathcal{L}(\alpha,\beta,\sigma|x,y,\sigma_x,\sigma_y),
\end{split}
\end{equation}
where $\beta'^3$ is the Jacobian of the transformation
$(\alpha,\beta)\mapsto(\alpha',\beta')$.
This works since for the priors we have $p(\alpha',\beta',\sigma) =
\abs{\beta^3}p(\alpha,\beta,\sigma)$ and for the likelihood we have
\begin{equation}
   \mathcal{L}(\alpha',\beta',\sigma|y,x,\sigma_y,\sigma_x) =
\mathcal{L}(\alpha,\beta,\sigma|x,y,\sigma_x,\sigma_y)
\end{equation}
as found above.

The result in this section
can be easily extended to cases where the
intrinsic scatter is not normally distributed or the measurement
errors are not distributed with a bivariate normal distribution, by
approximating the distributions with a weighted sum of Gaussians, as
used e.g.\ in the
\texttt{linmix\_err}-method of \citet{kelly2007}.
However, the result is not applicable for situations more general than
the linear case considered here, such as when the $M$ itself is not
trivially Euclidean, but includes e.g.\ angular or directional
measurements or when the relation $f$ is non-linear. In these cases, the
likelihood function and consequently the posterior distribution may have
to be evaluated numerically.

\subsubsection{Comparison to some existing approaches}\label{sc:comparison}

The equations \eqref{eq:llh-single}, \eqref{eq:ab-delta} and \eqref{eq:ab-sigma}
are a fundamentally symmetric way to describe the likelihood of parameters
$(\alpha, \beta, \sigma)$ specifying a line with orthogonal intrinsic scatter.
Several earlier works have incorporated intrinsic scatter
as an error parameter, added in quadrature to the
measurement errors along some specific measurement axis, as in e.g.\ the least likelihood method in
\citet{gultekin2009} and the \fitexy{} method \citep{press1992} as modified in
\citet{tremaine2002}.
These methods introduce intrinsic scatter into the measurement errors
of the dependent variable (for now taken to be $y$),
yielding a total variance of the form
\begin{equation}\label{eq:oldsigma}
    \bar{\sigma}^2 = \sigma_y^2 + \beta^2\sigma_x^2 + \sigma_{\text{int},y}^2,
\end{equation}
in the case of normally distributed intrinsic scatter in the $y$-direction,
where $\sigma^2_{\text{int},y}$ is the variance of the intrinsic scatter and
$\rho=0$ is assumed. This is not equivalent to equation~\eqref{eq:ab-sigma},
and in particular, the missing normalization factor $1+\beta^2$ leads to
$\bar{\sigma}\fromto\infty$ as $\beta\fromto\infty$. 
\add{We will now discuss how this difference arises.}

Firstly, assume that the underlying distribution of the data is a normal
distribution orthogonal to a line $y=\alpha+\beta x$, i.e.\ given by
equation~\eqref{eq:intgaussian} with $\ortho = (y-\alpha-\beta
x)/\sqrt{1+\beta^2}$. In this case the conditional distributions of the intrinsic
scatter in the $x$- and $y$-directions 
are also normal, with variances
\begin{align}
    \sigma^2_{\text{int},x} &= \frac{1+\beta^2}{\beta^2}\sigma^2 \label{eq:intsigmax} \\
    \sigma^2_{\text{int},y} &= (1+\beta^2)\sigma^2. \label{eq:intsigmay}
\end{align}
However, the marginal distributions of $x$ and $y$
are \emph{not} normal.
Now, the result in equation~\eqref{eq:oldsigma} can be
obtained from the likelihood produced by a single measurement,
\begin{equation}\label{eq:llh_integral}
\begin{split}
    &\mathcal{L}(\alpha,\beta|x_0,y_0,\sigma_x,\sigma_y,\sigma) = \\
    &\quad \iint_{\fR^2} h(x,y;x_0,y_0,\sigma_x,\sigma_y,\rho)\,\intdist(\ortho;\sigma)\,\ud x\,\ud y
\end{split}
\end{equation}
where $h$ is the 2-dimensional Gaussian given by
equation~\eqref{eq:2dgaussian} and $\intdist$ is the normal distribution,
equation~\eqref{eq:intgaussian}. \add{The assumptions yielding
equation~\eqref{eq:oldsigma} are to take} 
$\ortho = y-\alpha-\beta x$, i.e.\ distance
from the regression line \emph{in the $y$-direction}, and to set
$\sigma=\sigma_{\text{int},y}$.
\add{This yields} a normal distribution, as in equation~\eqref{eq:llh-single},
but with
\begin{align}
    \label{eq:ortho_a}
    \ortho^2 &= \ortho^2_a = (y_0-\alpha-\beta x_0)^2 \\
    \label{eq:sigma_a}
    \tilde{\sigma}^2 &= \tilde{\sigma}^2_a = \beta^2\sigma_x^2 + \sigma_y^2 -
    2\beta\rho\sigma_x\sigma_y + \sigma^2_{\text{int},y},
\end{align}
where the subscript $a$ refers to asymmetric, which is a point we will discuss
below. If we set $\rho=0$, we have the likelihood used in \citet{gultekin2009}, 
\begin{equation}\label{eq:gultekin-llh}
    \mathcal{L} = \frac{1}{\sqrt{2\pi\bar{\sigma}^2}} \exp\left(-\chi^2\right) 
                = \frac{
        \exp\left[
            -\frac{(y_0-\alpha-\beta x_0)^2}{2(\sigma_y^2 + \beta^2\sigma_x^2 + \sigma_{\text{int},y}^2)}
        \right]
    }{\sqrt{2\pi(\sigma_y^2 + \beta^2\sigma_x^2 + \sigma_{\text{int},y}^2)}},
\end{equation}
as well as the $\chi^2$ value used in \citet{tremaine2002}.
Likewise, if we instead use $\ortho = (y-\alpha-\beta x)/\sqrt{1+\beta^2}$, i.e.\ orthogonal
distance from the regression line, in the equation
\eqref{eq:llh_integral}, the result is the set of equations
\eqref{eq:llh-single}-\eqref{eq:ab-sigma}.
However, at this point we find that
making the obvious substitution from equation~\eqref{eq:intsigmay}
does \emph{not} make the set of equations
\eqref{eq:llh-single}-\eqref{eq:ab-sigma} equivalent to equation \eqref{eq:llh-single}
combined with the substitutions \eqref{eq:ortho_a} and \eqref{eq:sigma_a}.
\add{Furthermore the $\chi^2$ in
equation~\eqref{eq:gultekin-llh} is invariant under the
switch of independent and dependent coordinates, together with
$\sigma_{\text{int},y}\leftrightarrow\sigma_{\text{int},x}$.
However, the variance $\tilde{\sigma}^2_a$ and consequently the
likelihood are manifestly not invariant.
The equation~\eqref{eq:gultekin-llh} will give smaller likelihoods than
equation~\eqref{eq:llh-single} for $\beta \gg 1$ or $\beta\sim 0$,
depending on which variable is taken as the dependent one.
}


The fundamental reason for this state of affairs is the fact that if the
intrinsic scatter is modelled as a proper probability distribution in any
particular coordinate direction, the normalization of the intrinsic
scatter as a distribution orthogonal to the relation will change
in normalization with change in $\beta$. This reflects the breaking of the
rotational symmetry of the problem.
For example, assume that the intrinsic scatter is modelled as a
normal distribution in the $y$-direction, with variance
$\sigma^2_{\text{int},y}$. In this case the distribution of the
data, being the intrinsic scatter in a direction orthogonal to the
regression line, will need to have a normalization constant that goes
down with increasing slope $\beta$. Indeed, as $\beta\fromto\infty$, the
normalization constant will need to tend towards zero, to keep the
conditional probability distribution in the $y$-direction normalized to unity. This
leads to a vanishing likelihood for a vertical line
$\beta\fromto\infty$, which results from the fact that in this case
$\tilde{\sigma}_a\fromto\infty$ and consequently the resulting
likelihood
$\mathcal{L}(\alpha,\beta\fromto\infty|\mathcal{D})\fromto 0$.
In mathematical terms, if we have a distribution of the data
depending only on the orthogonal distance, $\intdist(\ortho)$, and we
have demanded that in $y$-direction we should have a proper probability
density function $p_{\text{int},y}(y)$, that is
\begin{equation}
    \int_{-\infty}^\infty p_{\text{int},y}(y)\,\ud y = \int_{-\infty}^{\infty} \intdist(\ortho(x(y),y))\,\ud y = 1,
\end{equation}
then necessarily
\begin{equation}
        \int_{-\infty}^{\infty} \intdist(\ortho)\,\ud\ortho =
        \int_{-\infty}^{\infty} \intdist(\ortho(x(y),y))
        \frac{\ud \ortho}{\ud y}\,\ud y 
        = \frac{1}{\sqrt{1+\beta^2}},
\end{equation}
which goes to zero as $\beta\fromto\infty$.
This is inconsistent, if we expect the distribution of the data,
$\intdist$, to have the same normalization no matter which way the
regression line might point.
Armed with this knowledge,
if we now multiply the right side of equation~\eqref{eq:gultekin-llh} by
$\sqrt{1+\beta^2}$ and substitute
$\sigma^2_{\text{int},y}$ from equation~\eqref{eq:intsigmay}, we \emph{do} arrive
at the equations \eqref{eq:llh-single}-\eqref{eq:ab-sigma}.
Based on this analysis, the likelihood of equation~\eqref{eq:llh-single}
is recommended for the purpose of fitting a regression
line with normally distributed intrinsic scatter to data with normally
distributed measurement errors. This point is also raised in
\citet{robotham2015}.

\begin{figure}
    \includegraphics[width=0.5\textwidth]{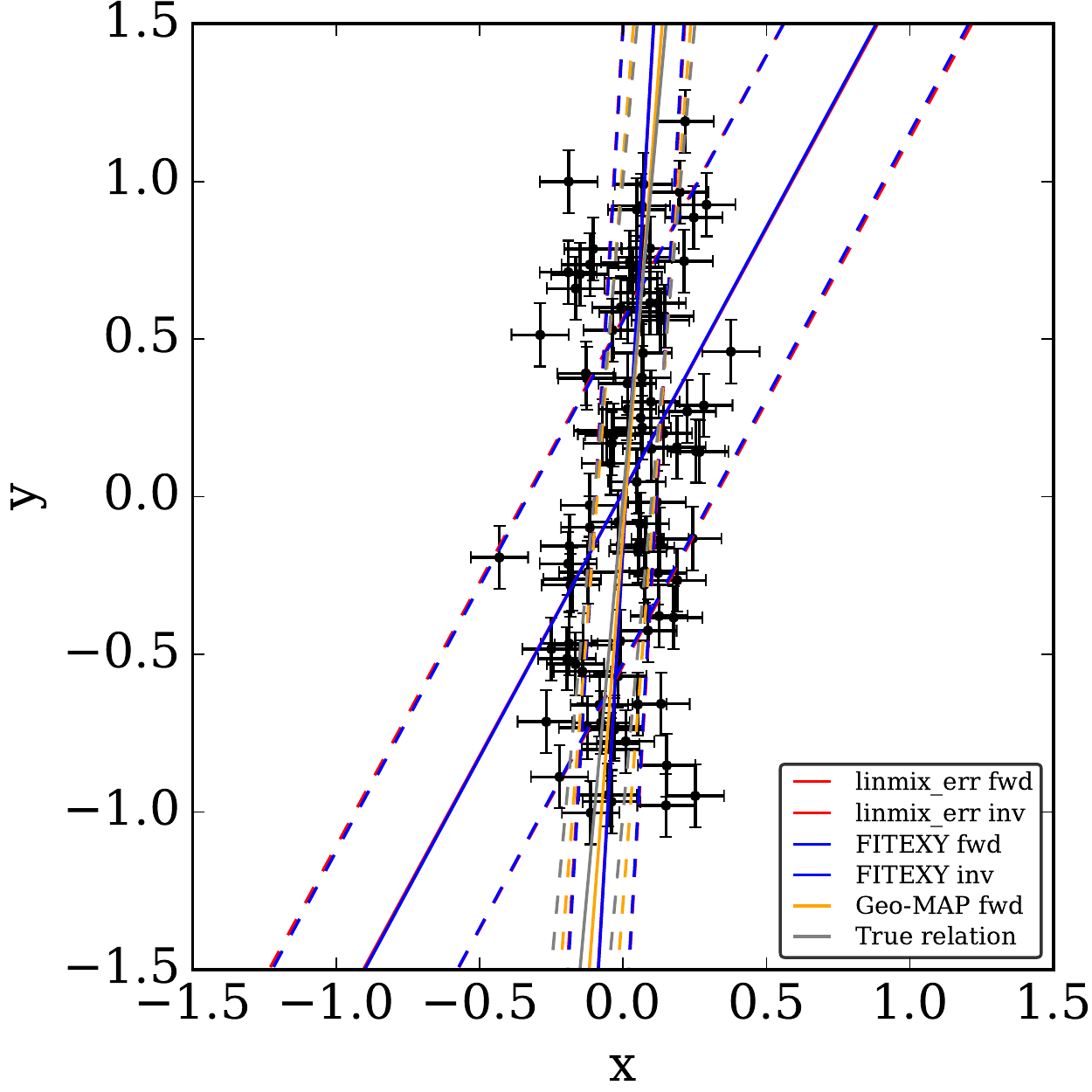}
    \caption{%
        Fits to simulated measurements with measurement errors
        $\sigma_x=\sigma_y=0.1$, sampled from a linear relation $y =
        10x$ with intrinsic scatter $\sigma=0.1$.
        Solid lines show the fitted relations and dashed lines
        indicate 1-$\sigma$ intrinsic scatter estimates.}
    \label{fig:comparison-plot}
\end{figure}

To investigate this conclusion numerically, we generated sets of simulated
observations from linear relations with intrinsic scatter.
The data were then fit using \linmixerr, \fitexy,\footnote{We used the
implementation available in the
\texttt{MPFIT} package \citep{markwardt2009} through the
\texttt{MPFITEXY} wrapper routine \citep{williams2010}.}
and by maximizing the posterior probability,
equation~\eqref{eq:2d-post}. This last method is referred to as Geo-MAP
(Geometric Maximum A Posteriori) hereafter. 
The parameter estimates and 1-$\sigma$ uncertainties for the \linmixerr{} method
were taken using the median and standard deviation of the Markov chain outputs. For the
\fitexy{} method the estimate and uncertainty provided by the algorithm were
used. For the Geo-MAP method a bootstrapping method was used together with a
numerical maximization of the posterior distribution to yield the parameter
estimates and 1-$\sigma$ uncertainties.
The simulated datasets were generated
assuming a relation $y = \alpha + \beta x$, with a normally distributed orthogonal intrinsic
scatter $\sigma=0.1$ and normally distributed measurement error with
$\sigma_x = \sigma_y = 0.1$ and $\rho=0$. 
In total six datasets were
generated, with $n_d=100$ data points each and with $\alpha\in\{0,10\}$
and $\beta\in\{0.1,1,10\}$. The datasets represent a situation where the intrinsic
scatter and measurement errors contribute equally to the observed
scatter, and as such approximations assuming the relative smallness of either are
maximally violated.
The data were generated with a uniform distribution $U(-1,1)$ along the
relation, and then shifted first according to $\sigma$ and then
according to $\sigma_x$ and $\sigma_y$.
The data was then fit both in the
forward direction, with $y = \alpha + \beta x$ as well as the inverse
direction, with $x = \alpha' + \beta' y$. For the inverse direction, the slope
$\beta = 1/\beta'$ was computed after the fit.
The slopes of the resulting fits are listed in
Table~\ref{tb:comparison-fits}. Figure~\ref{fig:comparison-plot}
illustrates the differences between the
methods for the $\alpha=0$, $\beta=10$ case.

The results show that for values of $\beta \gg 1$ the inverse fits with the
\linmixerr{} and \fitexy{} methods agree better with the data, albeit with high
uncertainties. Similarly, for $\beta \ll 1$ the forward fits give a better
agreement with the data. This is in line with the observations above, showing
that these methods give results tending towards lower values of $\beta$ for $\beta \gg 1$
when used in the forward direction,
and correspondingly towards higher values of $\beta$ for $\beta \sim 0$, when
used in the inverse direction.
It should be noted that in addition to yielding estimates of $\beta$ that are
much too high or low in these cases, both \linmixerr{} and \fitexy{} methods
give corresponding uncertainty estimates that are small, so that the true value of $\beta$
ends up as a multiple-$\sigma$ outlier.
Similarly as expected, the Geo-MAP
method gives nearly identical results for forward and inverse fits, up to the noise
caused by the bootstrapping procedure. In addition, the true value of $\beta$ is
always contained within the 1-$\sigma$ bounds except for the case of
$(\alpha,\beta)=(10,0.1)$. Finally, we note that none of the methods appear
invariant with respect to a shift with $\alpha$.

\begin{table}
    \caption{%
        Slopes of linear relations fit with the maximum a posteriori
        estimate of equation \eqref{eq:2d-post} (MAP)
        and the \linmixerr{} and \fitexy{} methods.
    }
    \label{tb:comparison-fits}
    \begin{tabularx}{0.5\textwidth}{lccc}
        \hline\hline

        Slope of data & $0.1$ & $1$ & $10$ \\
        Method & \multicolumn{3}{c}{Slope of fit $\pm$ 1-$\sigma$ uncertainties} \\ \hline

               & \multicolumn{3}{c}{$\alpha=0$} \\ \hline

        \linmixerr{} fwd & $0.09 \pm 0.02$ & $0.99 \pm 0.04$ & $1.67 \pm 0.70$ \\
        \linmixerr{} inv & $0.42 \pm 0.12$ & $1.08 \pm 0.05$ & $15.4 \pm 5.99$ \\
        \fitexy{} fwd    & $0.09 \pm 0.02$ & $0.99 \pm 0.04$ & $1.68 \pm 0.39$ \\
        \fitexy{} inv    & $0.41 \pm 0.06$ & $1.08 \pm 0.04$ & $15.5 \pm 6.10$ \\
        Geo-MAP   fwd    & $0.10 \pm 0.03$ & $1.04 \pm 0.05$ & $11.8 \pm 2.80$ \\
        Geo-MAP   inv    & $0.10 \pm 0.03$ & $1.04 \pm 0.04$ & $11.8 \pm 2.79$ \\

                & \multicolumn{3}{c}{$\alpha=10$} \\ \hline
        \linmixerr{} fwd & $0.04 \pm 0.02$ & $0.93 \pm 0.04$ & $1.79 \pm 0.57$ \\
        \linmixerr{} inv & $0.51 \pm 0.27$ & $1.01 \pm 0.04$ & $10.9 \pm 3.31$ \\
        \fitexy{} fwd    & $0.04 \pm 0.02$ & $0.93 \pm 0.04$ & $1.80 \pm 0.35$ \\
        \fitexy{} inv    & $0.52 \pm 0.13$ & $1.01 \pm 0.04$ & $10.9 \pm 3.24$ \\
        Geo-MAP   fwd    & $0.04 \pm 0.02$ & $0.98 \pm 0.04$ & $8.66 \pm 1.68$ \\
        Geo-MAP   inv    & $0.04 \pm 0.02$ & $0.98 \pm 0.04$ & $8.66 \pm 1.68$ \\

        \hline

    \end{tabularx}
    \parnotes
\end{table}

We conclude that methods based on likelihood examining distances
from the regression line in the direction of a particular coordinate
may give estimates of slope that are far from the true value while
simultaneously providing tight error bounds.
In particular, if the distance is
measured along the dependent variable, the resulting method will
be biased towards $\beta = 0$. \add{For methods, like the
maximum-likelihood method in \citep{gultekin2009},} this effect
remains even when the intrinsic scatter $\sigma\fromto 0$, in which case
the distribution $\intdist$ becomes a delta-function ridge, and the
problem reduces to Ordinary Least Squares (OLS) regression.
\add{This tendency towards lower slopes was reported} in \citet{park2012}
for all the methods used in the paper, namely OLS, BCES
\citep{akritas1996}, \fitexy{}, \linmixerr{}, and the
\citet{gultekin2009} maximum-likelihood method. Our results also agree
with \citet{isobe1990}, who studied different regression methods in the
limit of vanishing measurement error.

\subsection{Ellipses in Euclidean space}

\add{Next, we consider a case where the relation $f$ is properly
non-linear. With astrophysical applications in mind, we seek analytic
results for ellipses embedded in a Euclidean space. A possible application
could be, for example, to model the orbit of a stream of low mass
particles.
As such, for step
\ref{todo:pt1} we consider our measurements to be points in $M=\fR^3$
with the Euclidean metric. For step \ref{todo:pt2}, we define the
ellipse with
\begin{equation}
    f(\vct{x};\vct{\relpars}) = 
    \frac{\tilde{x}_1^2}{a^2} + \frac{\tilde{x}_2^2}{(1-e^2)a^2} - 1 = 0,
\end{equation}
where $\vct{\relpars} = (\vct{p},a,e,i,\omega,\Omega)$.
Here the parameters $\vct{\relpars}$ constitute $\vct{p}\in\fR^3$,
which specifies the position of the centre of the ellipse, and
$a$, $e$, $i$, $\omega$ and $\Omega$, which are the orbital elements, namely
semimajor axis, eccentricity, inclination, argument of pericentre and
longitude of the ascending node, respectively. 
The tilde signifies a coordinate transformation to a vector 
$\tilde{\vct{x}} = \mat{R}^T(\vct{x}-\vct{p})$, where the
matrix $\mat{R}$ rotates the Euclidean coordinate triad
$\{\hat{\vct{e}}_1, \hat{\vct{e}}_2, \hat{\vct{e}}_3\}$ into a coordinate
triad $\{\tilde{\vct{e}}_1, \tilde{\vct{e}}_2, \tilde{\vct{e}}_3\}$
aligned with the ellipse. That is, $\tilde{\vct{e}}_1$ points along the
semimajor and $\tilde{\vct{e}}_2$ along the semiminor axis.
The matrix $\mat{R}$ can be written explicitly as the product
\begin{equation}
    \begin{split}
    \mat{R} &= 
    \begin{pmatrix}
        \cos\Omega & -\sin\Omega & 0 \\
        \sin\Omega & \cos\Omega & 0 \\
        0 & 0 & 1
    \end{pmatrix}
    \begin{pmatrix}
        0 & 0 & 1 \\
        0 & \cos i& -\sin i \\
        0 & \sin i& \cos i
    \end{pmatrix}
    \\
    &\quad\times
    \begin{pmatrix}
        \cos\omega & -\sin\omega & 0 \\
        \sin\omega & \cos\omega & 0 \\
        0 & 0 & 1
    \end{pmatrix}.
    \end{split}
\end{equation}
In the ellipse-aligned frame, the ellipse can now be characterized also
as the curve
\begin{equation}
    \vct{g}(E) = a \cos E \tilde{\vct{e}}_1 + \sqrt{1-e^2}a \sin E
    \tilde{\vct{e}}_2,
\end{equation}
where $E\in[0,2\pi)$ is typically called the eccentric anomaly.
To use this convenient form, we will work in the coordinate frame
aligned with ellipse in the following.
}

\add{
We can now define the intrinsic coordinates (step \ref{todo:pt3}) by
defining $\para$ to be the arc length of the ellipse, with $\tau=0$ in
the direction of $\tilde{\vct{e}}_1$ and increasing towards
$\tilde{\vct{e}}_2$. We will need the relation between the eccentric
anomaly $E$ and the arc length $\tau$, given by
\begin{equation}
    \tau(E) = \sqrt{1-e^2} \, a\, \ellE(E, -e^2(1-e^2)^{-1}),
\end{equation}
where 
\begin{equation}
    \ellE(\phi,m) = \int_{0}^{\phi}\sqrt{1 - m\sin^2\theta}\,\ud\theta
\end{equation}
is the incomplete elliptic integral of the second kind. The
orthogonal coordinate axes can be defined so that $\hat{\vct{\ortho}}_1$ is
parallel to $\tilde{\vct{e}}_3$ and $\hat{\vct{\ortho}}_2$ points orthogonally
away from the ellipse, and thus can be given as 
\begin{equation}
    \begin{split}
        \hat{\vct{\ortho}}_2 
        &\propto \frac{2 \tilde{x}_1}{a^2} \tilde{\vct{e}}_1 + \frac{2\tilde{x}_2}{(1-e^2)a^2} \tilde{\vct{e}}_2 \\
        &= \frac{2 \cos E}{a} \tilde{\vct{e}}_1 + \frac{2 \sin E}{\sqrt{1-e^2}a} \tilde{\vct{e}}_2,
    \end{split}
\end{equation}
with normalization omitted for clarity.
}

\add{
For step \ref{todo:pt4}, we can now construct the coordinate
transformation $\crdchart\circ\surfchart^{-1}$ to find
\begin{equation}
    \tilde{\vct{x}}(\tau,\ortho_1,\ortho_2) =
    \vct{g}(E(\tau)) + \ortho_1\tilde{\vct{e}}_3 
    + \ortho_2\left(
\frac{2 \cos E}{a} \tilde{\vct{e}}_1 + \frac{2 \sin E}{\sqrt{1-e^2}a} \tilde{\vct{e}}_2
    \right).
\end{equation}
However, the inverse transformation is already very difficult to find
analytically, and further problems caused by the nonlinearity will now
also manifest.
}

\add{
At this point, step~\ref{todo:pt5}, we should specify the intrinsic
scatter model. To mitigate analytic problems, we would again like to use
the simplest normal distribution, so that 
\begin{equation}
    \intdist(\para, \ortho_1, \ortho_2;\sigma) 
    =
    \frac{\exp\left[-\frac{1}{2\sigma^2}(\ortho_1^2+\ortho_2^2) \right]}{2\pi\sigma^2}.
\end{equation}
However, we find that even if we consider a similarly simplified
measurement error distribution,
\begin{equation}
    h(\tilde{\vct{x}};\tilde{\vct{m}},\sigma_m) = 
    \frac{
        \exp\left[
        -\frac{1}{2\sigma_m^2}
        (\tilde{\vct{x}} - \tilde{\vct{m}})^T
        (\tilde{\vct{x}} - \tilde{\vct{m}})
        \right]
        }{(2\pi\sigma_m^2)^{3/2}},
\end{equation}
it is very difficult to obtain a closed form result
for the likelihood integral, equation~\eqref{eq:likelihood}.  Neither is
it obvious how to compute the form of $\intdist$ in the measurement
coordinates $\crdchart$, whether in the original or ellipse-aligned
form.
}

\add{
To make progress, we have to make some assumptions. The main assumption
we need is that the intrinsic scatter $\sigma$ is small compared to the
minimum radius of curvature of the ellipse, or $\sigma \ll (1-e^2)b$.
If this is the case, we can approximate
the intrinsic scatter distribution with a convolution of the ellipse and
a normal distribution, which in the ellipse-aligned coordinate chart yields
\begin{equation}
\begin{split}
    \intdist(\tilde{\vct{x}}) \sim
        \int_{\fR^3}
        &\delta\left(
            \frac{\tilde{y}_1^2}{a^2} + \frac{\tilde{y}_2^2}{(1-e^2)a^2} - 1
        \right) 
        \\
        &\quad \times \frac{
            \exp\left[
            -\frac{1}{2\sigma^2}
            (\tilde{\vct{y}} - \tilde{\vct{x}})^T
            (\tilde{\vct{y}} - \tilde{\vct{x}})
            \right]
        }{
            (2\pi\sigma^2)^{3/2}
        }\,\ud\tilde{\vct{y}},
\end{split}
\end{equation}
where $\delta$ is the Dirac delta distribution.
While this integral also resists evaluation in closed form, we can
approximate it by a mixture of normal distributions, by distributing individual
normal distributions on points equidistantly spaced in arc length.
The
number $N$ of component distributions required to make the distribution
smooth depends
on the intrinsic scatter $\sigma$. A Euclidean separation of less than
$\sigma$ is sufficient, and we can thus set
\begin{equation}
N = \left\lceil \frac{\tau(E=2\pi)}{\sigma} \right\rceil,
\end{equation}
where the brackets denote the ceiling function.
We can now complete step~\ref{todo:pt6}, and evaluate the likelihood
for a single measurement, yielding
\begin{equation}\label{eq:ellipsellh}
\begin{split}
    &\mathcal{L}(\vct{\relpars},\sigma|\tilde{\vct{m}},\sigma_m) =\\
    &\int_{\fR^3} 
    \frac{
        \exp\left[
        -\frac{1}{2\sigma_m^2}
        (\tilde{\vct{x}} - \tilde{\vct{m}})^T
        (\tilde{\vct{x}} - \tilde{\vct{m}})
        \right]
        }{(2\pi\sigma_m^2)^{3/2}}
    \\
    &\quad\quad \times
    \frac{1}{N} \sum_{i=1}^{N}
        \frac{
            \exp\left[
            -\frac{1}{2\sigma^2}
            (\tilde{\vct{x}} - \vct{g}(E_i))^T
            (\tilde{\vct{x}} - \vct{g}(E_i))
            \right]
        }{
            (2\pi\sigma^2)^{3/2}
        }
        \,\ud\tilde{\vct{x}} \\
    &\quad = 
    \frac{1}{N} \sum_{i=1}^{N}
    \frac{
        \exp\left[
            -\frac{1}{2(\sigma_m^2+\sigma^2)}
        (\vct{g}(E_i) - \tilde{\vct{m}})^T
        (\vct{g}(E_i) - \tilde{\vct{m}})
        \right]
        }{(2\pi[\sigma_m^2 + \sigma^2])^{3/2}},
\end{split}
\end{equation}
where the eccentric anomalies $E_i$ need to be numerically solved from
\begin{equation}
    \tau(E_i) = \tau(2\pi)\frac{i}{N}.
\end{equation}
The equation \eqref{eq:ellipsellh} is then straightforward to evaluate
numerically for each datapoint, completing step~\ref{todo:pt7}.
}

\subsection{Circles on a spherical surface}

\add{
Finally, we consider an example of a relation defined in a non-Euclidean
space. Specifically, we assume the measurements lie on $M=S^2$, the
two-dimensional spherical shell. We use the standard spherical coordinate chart
$(\theta,\phi)\in\fR^2$, for which the metric is given by the matrix
\begin{equation}
    \mat{G} = \begin{pmatrix} 1 & 0 \\ 0 & \sin^2\theta \end{pmatrix}.
\end{equation}
This choice is appropriate for positions on the plane of the sky, for example.
This definition completes step~\ref{todo:pt1}. For step~\ref{todo:pt2},
we choose a relation that defines a circle, by setting
\begin{equation}
    f(\theta, \phi; \vct{\relpars}) = \tilde{\theta} -
    \alpha_c = 0,
\end{equation}
where $\vct{\relpars}=(\theta_c, \phi_c, \alpha_c)$.
Here the parameters $(\theta_c, \phi_c)\in\fR^2$ specify the centre of
the circle, and $\alpha_c$ is equal to one-half of the angular size of
the circle. The tilde denotes a coordinate transformation which takes
the centre of the circle to the north pole, or $\tilde{\theta}_c = 0$.
This transformation is most conveniently realized by first transforming
the points $(\theta,\phi)$ to Cartesian three-vectors, using the
rotation matrix constructed in Section~\ref{sc:nd-case}, and then
converting back to spherical coordinates. The analytic form of the
transformation $(\theta,\phi)\mapsto(\tilde{\theta},\tilde{\phi})$ can
be given directly in terms of $(\theta,\phi)$ and $(\theta_c,\phi_c)$
but the result is unwieldy and omitted.
}

\add{
For steps \ref{todo:pt3} and \ref{todo:pt4} we now define the intrinsic
coordinates and the coordinate transformation between them and the
tilde-transformed coordinates.
A convenient choice is to set $\tilde{\phi} = \para$, so that the $\para$
coordinate increases with longitude as we go along the circle. We
likewise set $\tilde{\theta} = \alpha_c + \ortho$. This coordinate
transformation is multivalued in the sense that a given point
$(\tilde{\theta}, \tilde{\phi})$ corresponds to a infinite number of
intrinsic coordinate pairs, which can be written as
\begin{equation}
\begin{split}
    \surfchart\circ\crdchart^{-1}(\tilde{\theta}, \tilde{\phi}) &= 
    \{ (\tilde{\theta}-\alpha_c + n\,2\pi, \tilde{\phi}) | n\in\fZ \} \\
    &\quad\quad \cup
    \{ (\tilde{\theta}+\alpha_c + k\,2\pi, (\tilde{\phi}+\pi)) |
    k\in\fZ \}.
\end{split}
\end{equation}
The second set of intrinsic coordinate pairs above corresponds to geodesics
originating from the point of the circle on the other side of the
sphere, at $\tilde{\phi}+\pi$ (modulo $2\pi$).
}

\add{
For the definition of the intrinsic scatter distribution in
step~\ref{todo:pt5} we again use the one-dimensional normal
distribution, equation~\eqref{eq:intgaussian}. At this point, it is
advantageous to express the intrinsic scatter distribution in the
original coordinate frame as well.
Since the coordinate transformation is now multivalued, we must proceed
as explained in Section~\ref{sc:intdist}, to get
\begin{equation}\label{eq:sphere-intdist}
\begin{split}
    &\intdist(\tilde{\theta};\alpha_c,\sigma) \\
    &=
    \frac{1}{\sqrt{2\pi\sigma^2}}
    \!\!
    \sum_{n=-\infty}^{\infty}
    \!
    \left[
        e^{
            -\frac{(\tilde{\theta}-\alpha_c+n\,2\pi)^2}{2\sigma^2}
        }
        + e^{
            -\frac{(\tilde{\theta}+\alpha_c+n\,2\pi)^2}{2\sigma^2}
        }
        \right] \\
    &=
    \frac{1}{\sqrt{2\pi\sigma^2}}
    \Biggl[
        e^{
            -\frac{(\tilde{\theta}-\alpha_c)^2}{2\sigma^2}
        }
        \vartheta\left(i\frac{\pi(\tilde{\theta}-\alpha_c)}{\sigma^2},e^{-\frac{2\pi}{\sigma^2}}\right)
        \\
        &\quad\quad\quad\quad
        +
        e^{
            -\frac{(\tilde{\theta}+\alpha_c)^2}{2\sigma^2}
        }
        \vartheta\left(i\frac{\pi(\tilde{\theta}+\alpha_c)}{\sigma^2},e^{-\frac{2\pi}{\sigma^2}}\right)
    \Biggr],
\end{split}
\end{equation}
where $\vartheta$ is a Jacobi theta function corresponding to the
definition
\begin{equation}
    \vartheta(z,q) = \sum_{n=-\infty}^{\infty} q^{n^2} e^{2 n i z},
\end{equation}
with $q,z\in\fC$, $|q|<1$. Note that there is no dependence on
$\tilde\phi$. From this form of the $\intdist$
distribution it is easy to appreciate the close relation to the wrapped
normal distribution defined on the circle, also defined through a Jacobi
theta function. As such, this example shows
how the wrapped distributions naturally arise from the intrinsic
coordinate formalism.
}

\add{
To proceed further, we need to make some assumptions of the
error distribution of the measurements, which we denote by pairs
$(\tilde{\theta}_m, \tilde{\phi}_m)$. 
An attractive choice would be one of the
spherical generalizations of the normal distribution, such as the Kent
distribution or the Von Mises--Fisher distribution. However, these
combined with the intrinsic scatter distribution
\eqref{eq:sphere-intdist} and the metric of $S^2$ do not seem to easily
yield closed form results for the likelihood in step~\ref{todo:pt6}.
Instead, we have to assume that the measurement error is very small,
and approximately given by a uniform distribution within
a small cell
$[\tilde{\theta}_m \pm \Delta_m/2,
\tilde{\phi}_m \pm \Delta_m/(2\sin\tilde{\theta}_m)] \subset S^2$, 
where $\Delta_m$ specifies the (small) magnitude of the error.
}

\add{
With the help of this admittedly severe assumption, we can proceed to
step~\ref{todo:pt6} and compute the
likelihood for a single measurement, finding
\begin{equation}\label{eq:sphere-llh}
\begin{split}
    &\mathcal{L}(\vct{\relpars}, \sigma| \tilde{\theta}_m,
    \tilde{\phi}_m, \Delta_m) \sim \\
    &\quad
    \intdist(\tilde{\theta}_m;\alpha_c,\sigma) 
    +
    \frac{1}{12}
    \Delta_m^2
    \cot(\tilde{\theta}_m)
    \dass{\derfrac{\intdist(\tilde{\theta};\alpha_c,\sigma) }{\tilde\theta}}{\tilde\theta=\tilde{\theta}_m}
\end{split}
\end{equation}
to second order in $\Delta_m$.
While the likelihood~\eqref{eq:sphere-llh} is complicated by the
presence of the theta functions and their derivatives, these can be
efficiently evaluated numerically, and it is again straightforward to
numerically compute the value of the likelihood for each datapoint to
complete step~\ref{todo:pt7}.
}

\add{
We can compare the result to some existing approaches, such as presented in
\citet{jupp1987} and \citet{fujiki2009}. The former method
is based on splines, and the latter is based on Euclidean approximation
of small spherical distances and least squares fitting. 
The method in \citet{jupp1987} can
in principle cope with arbitrarily large normally distributed
measurement errors, but cannot incorporate intrinsic scatter, limits or
truncation. In addition, since the result is always a spline, no
parameter estimation for a predetermined curve is possible.
The method in \citet{fujiki2009} is suitable for parameter estimation,
but otherwise shares the disadvantages of the \citet{jupp1987} method in
addition to not incorporating measurement errors at all. In this sense,
the geometric approach presented here compares favourably,
although it presents mathematical difficulties if curves other than circles
are to be fit.
}


\subsection{What is learned from the examples}

\add{
The examples clearly demonstrate that analytic results are only easy to
obtain in the case where $M$ is Euclidean and the relation to be fit is
linear, or more exactly a geodesic. In the last two examples,
simplifying assumptions are necessary to obtain a closed form for the
likelihood. As such, while the linear Euclidean case of normally
distributed orthogonal scatter is essentially solved here completely,
much future work is required in the context on non-linear relations and
non-Euclidean spaces.
}

\add{In addition, the choice of}
parameterization of the relation to be fit has some important consequences.
Firstly, the chosen parameterization directly affects the form of the prior
probability distribution, \add{as discussed in
Section~\ref{sc:intdist}}.
In addition, different parameterizations may not be numerically well-behaved
everywhere. For example, the $\vct{p}$-parameterization \add{used in
Sections \ref{sc:nd-case} and \ref{sc:n-k-case}}
will likely exhibit numerical
instability if \add{any of the relation hyperplanes} passes \st{very} close to the coordinate
origin, in which case $\norm{\vct{p}}\fromto 0$ and the orientation of the
plane becomes indeterminate. The $(\alpha, \beta)$-parameterization in
Section \ref{sc:lincase} suffers from
similar difficulties for nearly vertical lines, as then the values
$\beta\fromto\pm\infty$ become degenerate.

\section{The \msigma{} relation}\label{sc:msigma}

As a typical application, the methodology described above can be applied 
to the \msigma{} relation, typically written in the form
\begin{equation}\label{eq:msigma}
    \log_{10}\left(\frac{M_\text{BH}}{\Msun}\right)= \alpha + \beta
    \log_{10}\left(\frac{\sigma}{200\,\si{km.s^{-1}}}\right).
\end{equation}
The \msigma{} relation is an important correlation between the mass
$M_\text{BH}$ of the supermassive
black hole in the centre of a galaxy and the velocity dispersion $\sigma$ of the
galactic bulge.\footnote{%
To avoid confusion, the intrinsic scatter will be denoted with
$\sigma_\text{int}$ in this section.
}
Since the initial discovery of this correlation
(\citealt{ferrarese2000,gebhardt2000}, but see also \citealt{magorrian1998}), 
it has been re-established several
times, using both larger datasets and different statistical procedures
\citep[e.g.][]{tremaine2002,novak2006,gultekin2009,mcconnell2011,graham2011,beifiori2012,mcconnell2013,saglia2016,bosch2016}.
The existence of the \msigma{} relation and analogous relations, such as the
correlation $M_\text{BH}\text{--}M_\text{bulge}$ with the galaxy
bulge mass, have also been confirmed in numerical
simulations, both in galaxy merger simulations
\citep[e.g.][]{dimatteo2005,johansson2009a,johansson2009b,choi2014}
and in cosmological simulations
\citep[e.g.][]{sijacki2007,dimatteo2008,booth2009,sijacki2015}.
However, all of these studies have used statistical methods which treat
one observable as independent and the other observable as dependent.
Redoing the fit with the independent observable as the dependent and
vice versa produces a significantly different correlation slope, and in
some cases also affects the estimated value of the intrinsic scatter \citep{park2012}.
Consequently, there is some controversy in the astronomical literature
as to whether it is more
suitable to fit $M_\text{BH}$ as a function of $\sigma$ (forward regression, using the
definition in \citealt{park2012}), or the other way around (inverse regression,
respectively) in the presence of intrinsic scatter.
For a review of the debate, see \citet{graham2016} and the references therein.
That different slopes are produced by switching dependent and independent
variables has been known for a long time \citep[e.g.][]{pearson1901}, and
amounts to asking two different questions: what is the most likely value for
$M_\text{BH}$ given $\sigma$ or vice versa. However, it is equally well known
that if one is interested in the functional relation between the observables,
then a symmetric method should be used \citep[see e.g.][]{isobe1990}.

The approach in Section~\ref{sc:lincase} presents a symmetric Bayesian solution
to fitting a linear relation between two observables, incorporating
heteroscedastic errors, upper limits and intrinsic scatter. As such, we would
expect it to yield a non-biased estimate of the intrinsic slope $\beta$ of the
relation \eqref{eq:msigma}, regardless of which way the relation is fit.
To investigate this, the three datasets used
in Table~1 of \citet{park2012} were analysed. The data are originally from
\citet{gultekin2009}, \citet{mcconnell2011} and \citet{graham2011}.
In order to compare with the results in \citet{park2012}, the
measurement errors were modelled as uncorrelated bivariate Gaussians in
the logarithmic space, leaving out all upper limits.
Following \citet{park2012}, the standard deviations were set equal to mean errors, i.e.\ 
$\sigma_{\log_{10} M_\text{BH}} = (\log_{10} M_\text{BH,high}-\log_{10} M_\text{BH,low})/2$ and
similarly for the velocity dispersions. In addition, for the
\citet{graham2011} sample, which lacks velocity dispersion error data,
the velocity dispersion errors were set to 10\% and then propagated to
averaged logarithmic errors.
Finally, a dataset compiled from a multitude of sources
used in \citet{bosch2016} was used. For this dataset, the errors were used as
given, and upper limits were also incorporated in the fit.

For each dataset, the values for the same set of parameters as in
\citet{park2012} were computed: the intercept $\alpha$, slope $\beta$ and
intrinsic scatter along the $M_\text{BH}$-axis $\sigma_{\text{int},M_\text{BH}}$. 
The results were computed for forward regression, equation~\eqref{eq:msigma}, as well as the inverse
regression, converting back to equivalent forward values in the end.
The parameter values and uncertainties were
estimated using the posterior distribution, equation \eqref{eq:2d-post},
in two complementary ways.

The first set of estimates (Geo-MAP, hereafter),
were derived by numerically maximizing the posterior distribution
combined with a bootstrap resampling procedure.
For each dataset, a total of $n_d(\log n_d)^2$ bootstrap samples were
constructed \citep{babu1983,feigelson2012}, where $n_d$ is the number of data
points. The parameters were then estimated using the median parameter
values. Parameter uncertainties at 1-$\sigma$ level were estimated using
median absolute deviations scaled to correspond to standard deviations for a normal distribution.

For the second set of estimates (Geo-MCMC, hereafter),
the posterior distribution was sampled
with the Markov chain Monte Carlo (MCMC) sampler \texttt{emcee} \citep{emcee2013}. Convergence during
sampling was monitored with the potential scale reduction factor $R$
\citep{gelman2013}, and sampling was continued until $R<1.01$ was
achieved. The parameter estimates were \st{then} obtained as the values
corresponding to the sample with maximum posterior probability. The
parameter uncertainties at 1-$\sigma$ level were then estimated by 
constructing credible regions containing $0.6827$ of the posterior
probability mass. This was done by starting from the maximum posterior probability sample, and
descending in posterior probability until the limit was exceeded. The extent of the credible
region in each parameter direction was then used to compute the upper and lower 1-$\sigma$
limits.

\begin{figure*}
    \begin{tabular}{cc}
        \includegraphics[width=0.5\textwidth]{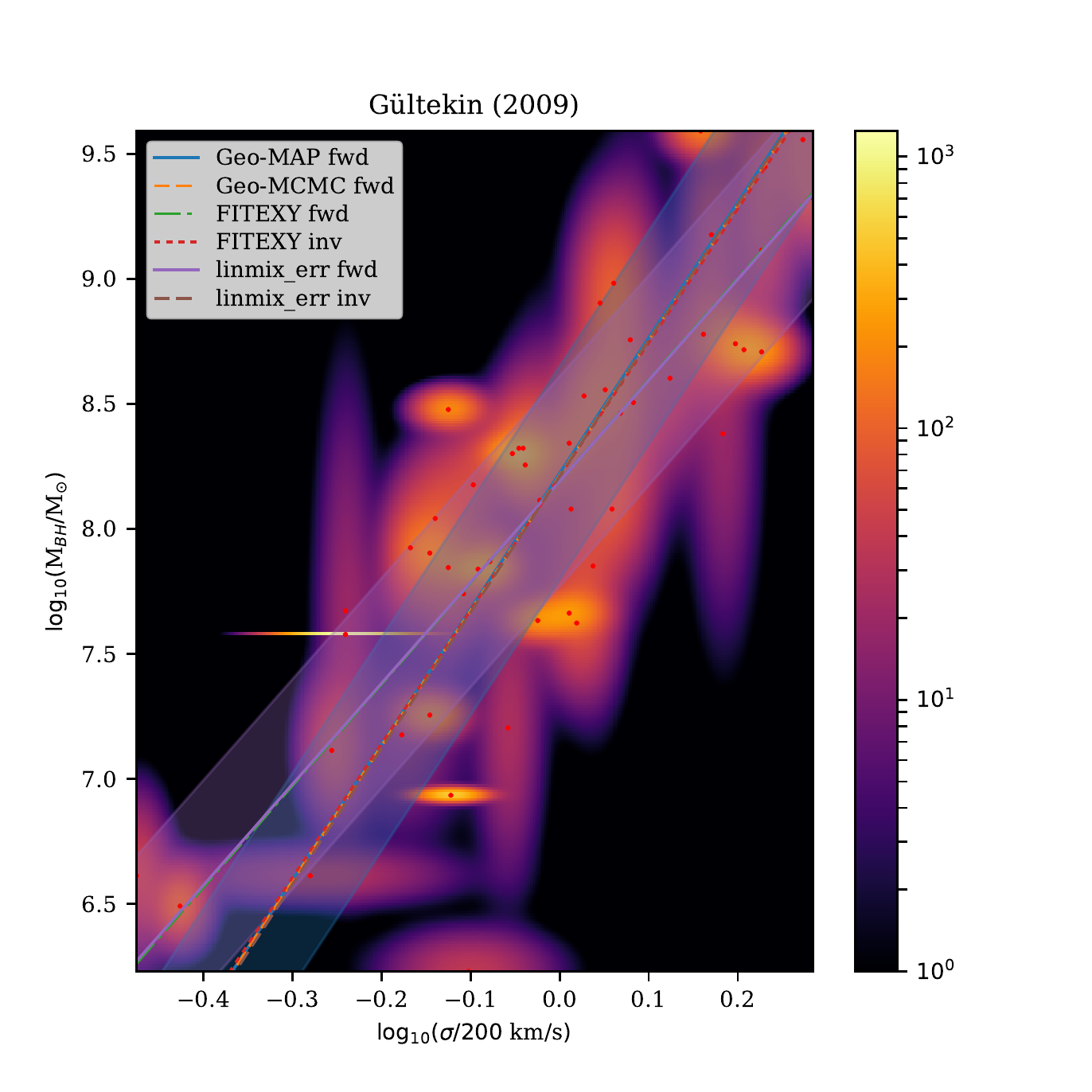} &
        \includegraphics[width=0.5\textwidth]{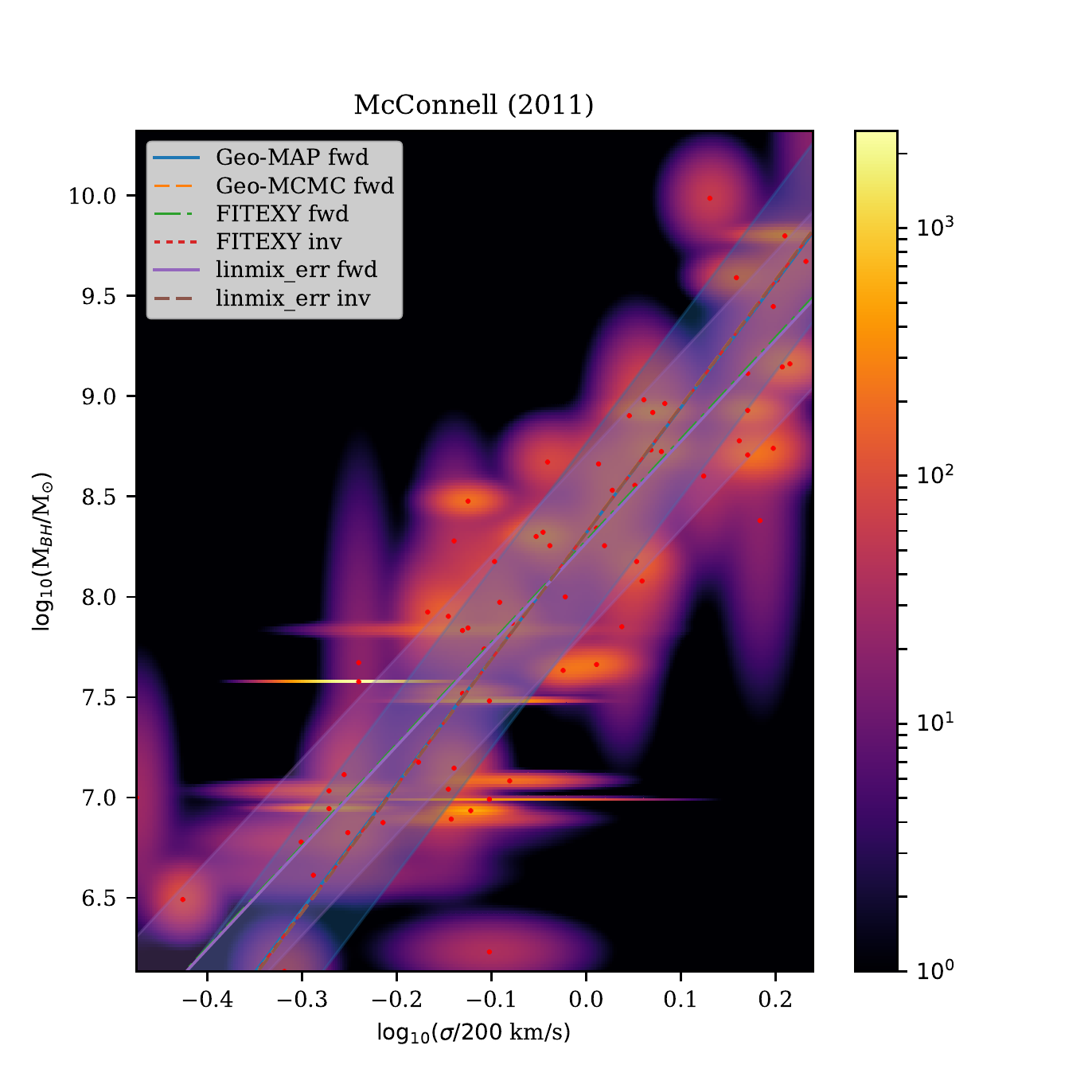} \\
        \includegraphics[width=0.5\textwidth]{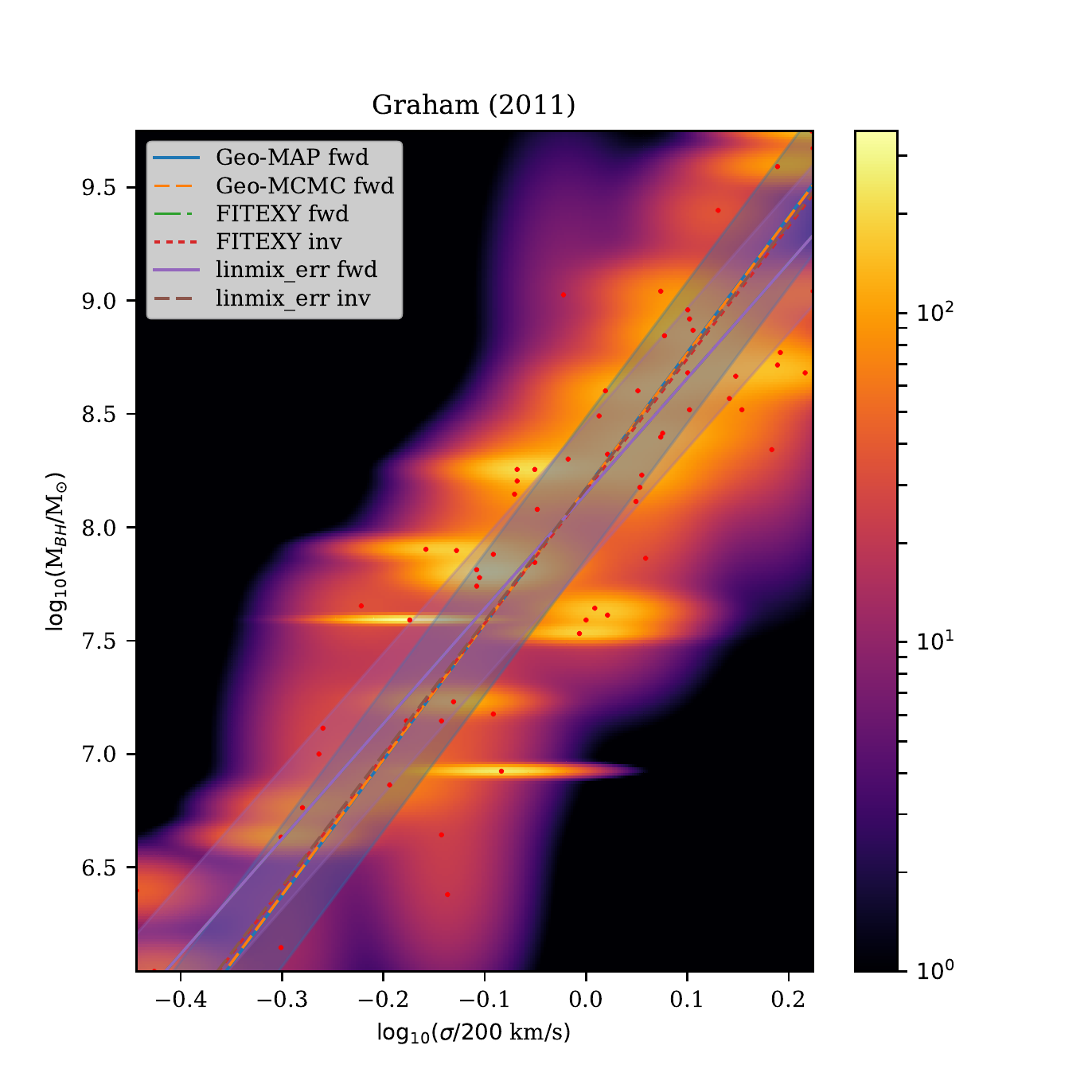} &
        \includegraphics[width=0.5\textwidth]{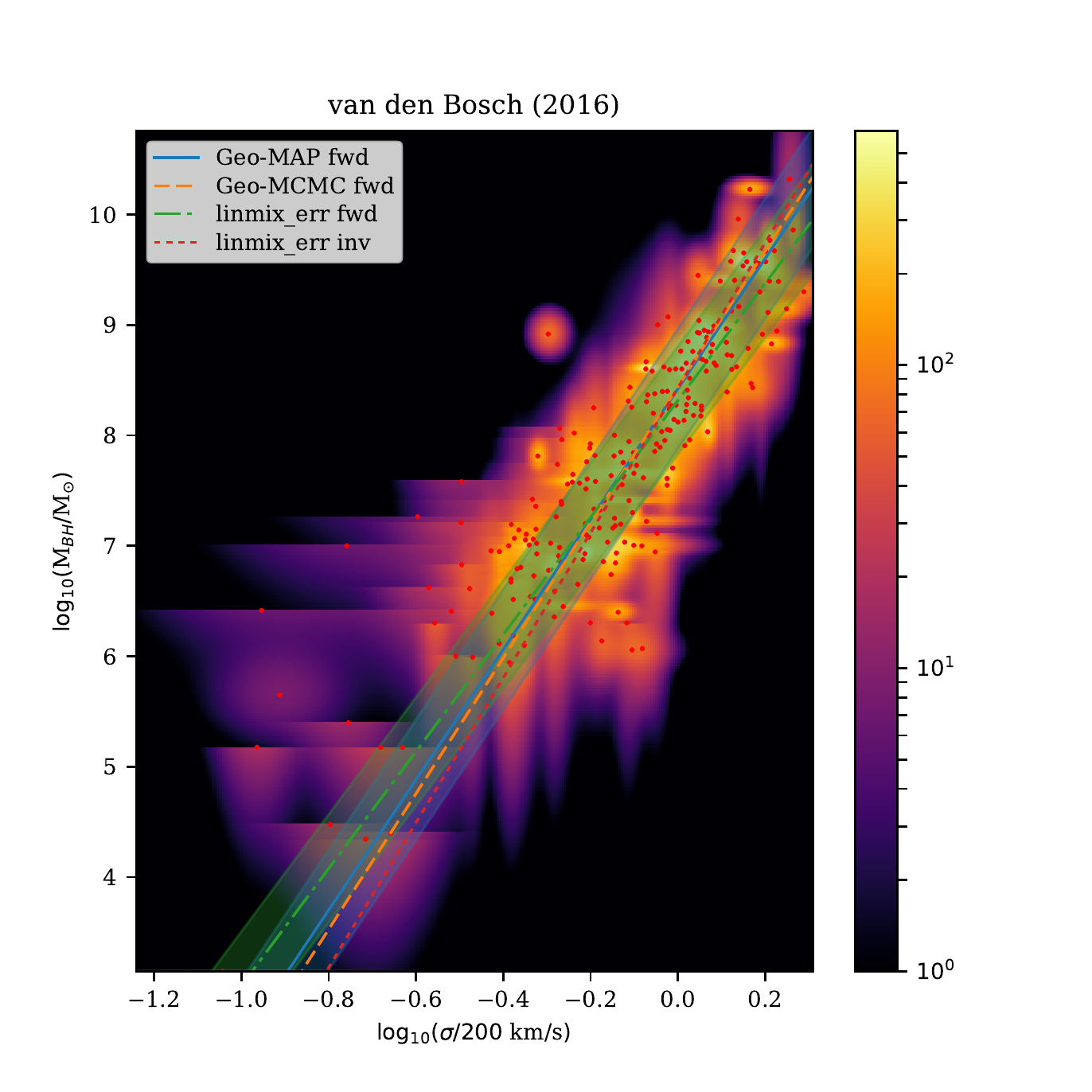} \\
    \end{tabular}
    \caption{
        \msigma{} relation
        $\log_{10}(M_\text{BH}/\Msun) = \alpha + \beta\log_{10}(\sigma/200\,\si{km.s^{-1}})$ 
        derived using the datasets in \citet{gultekin2009} (top left),
    \citet{mcconnell2011} (top right), \citet{graham2011} (bottom left),
and \citet{bosch2016} (bottom right). The datasets have been rendered
as density maps, assuming two-dimensional Gaussian probability densities
derived from the 1-$\sigma$ errors of the data points. The data points are shown
with red dots. The upper
limits in \citet{bosch2016} data are rendered as a product of
one-dimensional Gaussian and the logarithmic $y$ upper limit density,
equation \eqref{eq:loguppery}. The fitted relations are overplotted
with lines. Intrinsic scatter 1-$\sigma$ limits are shown with shaded
regions for the Geo-MAP and \linmixerr{} forward fits for demonstration.
The values of intrinsic scatter produced by the rest of the methods are comparable (see
Table~\ref{tb:park-comparison}).
\label{fig:park}
}
\end{figure*}

The results of this analysis are displayed in
Table~\ref{tb:park-comparison}.
Shown also are the results from
\citet{park2012} containing both forward and inverse
regressions with the \fitexy{} \citep{press1992,tremaine2002} 
and `Bayesian' (i.e.\ \linmixerr{}) methods.
In addition, fits for the \citet{bosch2016} dataset without upper limits were
computed separately for all methods.
A graphical representation of the datasets and the relations
obtained with the different methods is shown in Figure~\ref{fig:park}.

\begin{table*}
    \caption{%
        \msigma{} relations
        $\log_{10}(M_\text{BH}/\Msun) = \alpha +
        \beta\log_{10}(\sigma/200\,\si{km.s^{-1}})$ 
        derived using the datasets in \citet{gultekin2009},
        \citet{mcconnell2011}, \citet{graham2011} and \citet{bosch2016}, using
        methods \fitexy{}, \linmixerr{} and the Geometric method of this
        paper. For the Geometric method, results with both
        bootstrapped (Geo-MAP) and MCMC-sampled (Geo-MCMC) maximum a
        posteriori estimate are shown.
        The results for the \fitexy{} and \linmixerr{} methods for the first
        three datasets are from \citet{park2012}.
    }
    \label{tb:park-comparison}
    \begin{tabularx}{\textwidth}{XXXXX}
        \hline\hline
        Method & Intercept $\alpha$ & Slope $\beta$ & $\sigma_{\text{int},M_\text{BH}}$\tnote{a} & $\sigma_\text{int}$\tnote{a} \\ \hline

        \multicolumn{5}{c}{\citet{gultekin2009} data} \\ \hline

        \fitexy{} fwd    & $8.19\pm0.06$ & $4.06\pm0.32$ & $0.39\pm0.06$ & $0.093$ \\
        \fitexy{} inv    & $8.21\pm0.07$ & $5.35\pm0.66$ & $0.45\pm0.09$ & $0.083$ \\
        \linmixerr{} fwd & $8.19\pm0.07$ & $4.04\pm0.40$ & $0.42\pm0.05$ & $0.101$ \\
        \linmixerr{} inv & $8.21\pm0.08$ & $5.44\pm0.56$ & $0.49\pm0.09$ & $0.089$ \\
        Geo-MAP fwd      & $8.22\pm0.07$ & $5.42\pm0.72$ & $0.43\pm0.08$ & $0.076\pm0.008$ \\
        Geo-MCMC fwd     & $8.21^{+0.14}_{-0.14}$ & $5.41^{+1.10}_{-0.79}$ & $0.44^{+0.12}_{-0.09}$ & $0.080^{+0.022}_{-0.016}$ \\

        \hline
        \multicolumn{5}{c}{\citet{mcconnell2011} data} \\ \hline

        \fitexy{} fwd    & $8.28\pm0.06$ & $5.07\pm0.36$ & $0.43\pm0.05$ & $0.083$\\
        \fitexy{} inv    & $8.32\pm0.06$ & $6.29\pm0.49$ & $0.47\pm0.06$ & $0.073$\\
        \linmixerr{} fwd & $8.27\pm0.06$ & $5.06\pm0.36$ & $0.44\pm0.05$ & $0.085$\\
        \linmixerr{} inv & $8.32\pm0.07$ & $6.31\pm0.46$ & $0.49\pm0.07$ & $0.077$\\
        Geo-MAP fwd      & $8.32\pm0.05$ & $6.26\pm0.51$ & $0.45\pm0.06$ & $0.070\pm0.008$ \\
        Geo-MCMC fwd     & $8.32^{+0.13}_{-0.12}$ & $6.31^{+0.90}_{-0.71}$ & $0.46^{+0.11}_{-0.08}$ & $0.071^{+0.017}_{-0.013}$ \\

        \hline
        \multicolumn{5}{c}{\citet{graham2011} data} \\ \hline

        \fitexy{} fwd    & $8.15\pm0.05$ & $5.08\pm0.34$ & $0.31\pm0.04$ & $0.060$ \\
        \fitexy{} inv    & $8.16\pm0.05$ & $5.84\pm0.42$ & $0.33\pm0.05$ & $0.056$ \\
        \linmixerr{} fwd & $8.15\pm0.05$ & $5.08\pm0.36$ & $0.31\pm0.05$ & $0.060$ \\
        \linmixerr{} inv & $8.17\pm0.06$ & $5.85\pm0.42$ & $0.34\pm0.06$ & $0.057$ \\
        Geo-MAP fwd      & $8.17\pm0.04$ & $6.00\pm0.41$ & $0.31\pm0.05$ & $0.051\pm0.006$ \\
        Geo-MCMC fwd     & $8.17^{+0.11}_{-0.11}$ & $5.98^{+0.87}_{-0.67}$ & $0.32^{+0.11}_{-0.10}$ & $0.052^{+0.018}_{-0.016}$ \\

        \hline
        \multicolumn{5}{c}{\citet{bosch2016} data} \\ \hline

        \linmixerr{} fwd\tnote{b} & $8.32\pm0.04$ & $5.30\pm0.22$ & $0.49\pm0.03$ & $0.091$ \\
        Geo-MAP fwd               & $8.42\pm0.04$ & $5.90\pm0.26$ & $0.55\pm0.06$ & $0.091\pm0.008$ \\
        Geo-MCMC fwd     & $8.44^{+0.09}_{-0.09}$ & $6.15^{+0.42}_{-0.40}$ & $0.54^{+0.07}_{-0.06}$ & $0.087^{+0.011}_{-0.009}$ \\

        \hline
        \multicolumn{5}{c}{\citet{bosch2016} data, no upper limits} \\ \hline

        \fitexy{} fwd    & $8.35\pm0.04$ & $4.91\pm0.23$ & $0.48$\tnote{c} & $0.096$  \\
        \fitexy{} inv    & $8.42\pm0.05$ & $6.51\pm0.30$ & $0.55$\tnote{c} & $0.084$  \\
        \linmixerr{} fwd & $8.34\pm0.04$ & $4.92\pm0.25$ & $0.49\pm0.03$   & $0.098$  \\
        \linmixerr{} inv & $8.43\pm0.05$ & $6.57\pm0.32$ & $0.56\pm0.05$   & $0.084$  \\
        Geo-MAP fwd      & $8.43\pm0.04$ & $6.70\pm0.40$ & $0.54\pm0.07$   & $0.080\pm0.008$ \\
        Geo-MCMC fwd     & $8.43^{+0.10}_{-0.09}$ & $6.68^{+0.67}_{-0.55}$ & $0.55^{+0.08}_{-0.06}$ & $0.082^{+0.011}_{-0.010}$ \\
        \hline

    \end{tabularx}
    \begin{raggedright}
    \\[1em]
    \textbf{Notes.}\\
    \tnote{a} The orthogonal intrinsic scatter $\sigma_\text{int}$ has been
    converted to intrinsic scatter along the $M_\text{BH}$ coordinate via the
        equation \eqref{eq:intsigmay} for both Geo methods. 
        Similarly, the scatter in $M_\text{BH}$-direction,
        $\sigma_{\text{int},M_\text{BH}}$, has
        been converted to equivalent orthogonal intrinsic scatter via the same
        equation for the \linmixerr{} and \fitexy{} methods. \\

    \tnote{b} The \linmixerr{} method only supports upper limits on the
    dependent variable. As such, inverse fits cannot be computed. \\

    \tnote{c} The \fitexy{} method does not give error estimates for the intrinsic
        scatter. \\
    \end{raggedright}
\end{table*}

Table~\ref{tb:park-comparison} indicates that the results computed using the
geometrical approach presented in this paper are more consistent with the
inverse fits done using the \linmixerr{} and \fitexy{} algorithms. This is not
surprising in light of the discussion in the previous section, since the
slope of the \msigma{} relation is high and we expect the
forward fits to be biased towards low slopes in this situation. 
The effects of this bias can also
be seen from the best-fitting values of the orthogonal intrinsic scatter,
$\sigma_{\text{int}}$, for which the forward fits yield consistently higher
values than the inverse fits for \linmixerr{} and \fitexy{}. It seems that in
these methods the lower value for the slope is compensated by a higher estimate
for the intrinsic scatter. Note that this behaviour cannot be appreciated by
looking at the values of the scatter in $M_\text{BH}$-direction, $\sigma_{\text{int},M_\text{BH}}$,
since these are not truly intrinsic, but depend on $\beta$ and indeed show 
opposite behaviour. \add{In contrast, the geometric methods yield identical
results (up to sampling noise) in forward and inverse directions, so
only forward results are shown in Table~\ref{tb:park-comparison}.}
The geometric methods also give
consistently smaller estimates for the orthogonal intrinsic scatter. The
1-$\sigma$ errors for all parameters are in general comparable between the
methods, with the exception of the fully Bayesian MCMC approach, which
consistently gives more conservative 1-$\sigma$ errors.

Finally, it seems that if there indeed is a fundamental approximately linear
relation between the logarithms of the mass of a supermassive black hole mass
and the velocity dispersion of its host galaxy, \emph{the slope of the relation is
likely to be $\gtrsim 6$}, at least based on the \citet{bosch2016} data, which is
the most comprehensive dataset used here.
This is in contrast to the values around $4$--$5$ often obtained in the
literature (e.g.\ $4.8$, \citealt{ferrarese2000}; $3.75$, \citealt{gebhardt2000};
$4.02$, \citealt{tremaine2002}; $4.24$, \citealt{gultekin2009}; $5.64$,
\citealt{mcconnell2013}), including the value $5.35$ derived in
\citet{bosch2016}. These values were all derived using a method or a variation of a
method described in Section~\ref{sc:comparison}, fitting the \msigma{} relation
in the `forward' direction ($M_\text{BH}$ as a function of $\sigma$), in which it has a high
numerical value for the slope. 
This then causes the fitted value of the slope to be biased towards lower values.
However, as can be seen from Figure~\ref{fig:park}, the numerically rather
different estimates of the slope are not visually at all that obvious, with a
change in slope of $5$ to $6$ corresponding only to a $\sim \ang{1.8}$
change in angle of the regression line with respect to a fixed
direction, such as the $x$-axis.

The obtained slope also
crucially depends on the observational and other biases the data may
have. Indeed, it is evident from Table~\ref{tb:park-comparison}
that the choice of dataset has an effect that
is roughly comparable to the effect of the choice of method. 
It should also be noted that the fundamental relation, if there
is one, may involve more than two physical quantities or their measurable
proxies, as suggested in \citet{bosch2016}. Fitting relations to a sampled
projection of this hyperplane would then in general yield a higher value for the
intrinsic scatter and slope that is offset from the `true' value.
As such, there is an urgent need
for more high-quality data in order to say much with any certainty
regarding the
slope of the \msigma{} relation (or the possible multi-variable
generalizations), its possible evolution with redshift, or whether it truly is
linear across the entire range of black hole masses.

\section{Summary}\label{sc:summary}

We have presented a mathematical formalism for representing physical relations
as submanifolds $S$ of a Riemannian manifold of observables $M$.
In this geometric approach, intrinsic scatter in the relation can be
accommodated with probability distributions defined on the normal spaces of $S$.
Given a probability distribution of data, and \st{a} parameterizations of the
relation and the intrinsic scatter distribution, the formalism then yields a
Bayesian posterior probability for the parameters of the relation and the
intrinsic scatter distribution, equation~\eqref{eq:postdist}. The novelty of our
formulation is \st{in} that it fully accommodates arbitrary measurement errors, both left
and right censored data (upper and lower limits, respectively), truncation
(non-detections) and extends the concept of intrinsic scatter both to non-linear
relations and relations that define a submanifold of codimension greater than
one.

We have derived explicit analytic results for the likelihood and the posterior
distribution first in the case where the postulated relation defines a linear
$n-1$-dimensional hyperplane. We then extended this result to the case where the
relation defines an $n-k$-dimensional affine subspace, for an arbitrary $k$, a
result we believe to be potentially highly useful for seeking out the
most likely codimension of a correlation within a set of $n$-dimensional data.
Finally, we have derived the
likelihood and posterior distribution in the case of a line in two dimensions,
and discussed its implications at length. We also compared the results given by
our method with two established methods widely used in astronomical literature,
namely \fitexy{} and \linmixerr{}. We demonstrated that our inherently
symmetrical geometrical approach is preferable in situations where the
data obeys a relation with a slope much larger or smaller than one, and
measurement errors and intrinsic scatter are severe and equally important.

Finally, we used our method to fit the \msigma{} relation, between the mass
$M_\text{BH}$
of a supermassive black hole in a galactic bulge, and the stellar velocity dispersion
$\sigma$ of the bulge, using several
published datasets. We compared our results to the fits in the
literature, and find that our results support a slope of $\sim 6$, clearly higher
than the slopes $\sim4\text{--}5$ derived in the literature. We note that this
difference is mainly due to the methods used to derive the literature results.
We show that if these methods are used `in reverse', to fit $\sigma$ as a
function of $M_\text{BH}$, and then inverting the slope, the results are in much better
agreement with ours. This is due to the tendency of standard methods, 
such as \fitexy{} or \linmixerr{},
which do not respect the geometric symmetry of the problem, to misestimate
steep slopes in the presence of intrinsic scatter (see Section~\ref{sc:comparison} for
the discussion).

\section*{Acknowledgements}

The author is most grateful \add{to the anonymous referee for comments and
ideas that have contributed to a much improved manuscript, as well as} 
to Peter H.~Johansson for extensive comments on the
manuscript drafts.
The numerical computations in this work have benefited from the Python
libraries NumPy, SciPy and SymPy \citep{numpy, scipy, sympy}. The
Figures have been rendered with the help of the Python
library Matplotlib \citep{matplotlib}.
This work has made use of NASA's Astrophysics Data System Bibliographic
Services. The research for this publication was supported by the Academy of
Finland grant no.~1274931.




\bibliographystyle{mnras}
\bibliography{refs} 

\begin{thebibliography}{}
\makeatletter
\relax
\def\mn@urlcharsother{\let\do\@makeother \do\$\do\&\do\#\do\^\do\_\do\%\do\~}
\def\mn@doi{\begingroup\mn@urlcharsother \@ifnextchar [ {\mn@doi@}
  {\mn@doi@[]}}
\def\mn@doi@[#1]#2{\def\@tempa{#1}\ifx\@tempa\@empty \href
  {http://dx.doi.org/#2} {doi:#2}\else \href {http://dx.doi.org/#2} {#1}\fi
  \endgroup}
\def\mn@eprint#1#2{\mn@eprint@#1:#2::\@nil}
\def\mn@eprint@arXiv#1{\href {http://arxiv.org/abs/#1} {{\tt arXiv:#1}}}
\def\mn@eprint@dblp#1{\href {http://dblp.uni-trier.de/rec/bibtex/#1.xml}
  {dblp:#1}}
\def\mn@eprint@#1:#2:#3:#4\@nil{\def\@tempa {#1}\def\@tempb {#2}\def\@tempc
  {#3}\ifx \@tempc \@empty \let \@tempc \@tempb \let \@tempb \@tempa \fi \ifx
  \@tempb \@empty \def\@tempb {arXiv}\fi \@ifundefined
  {mn@eprint@\@tempb}{\@tempb:\@tempc}{\expandafter \expandafter \csname
  mn@eprint@\@tempb\endcsname \expandafter{\@tempc}}}

\bibitem[\protect\citeauthoryear{{Akritas} \& {Bershady}}{{Akritas} \&
  {Bershady}}{1996}]{akritas1996}
{Akritas} M.~G.,  {Bershady} M.~A.,  1996, \mn@doi [\apj] {10.1086/177901},
  \href {http://adsabs.harvard.edu/abs/1996ApJ...470..706A} {470, 706}

\bibitem[\protect\citeauthoryear{Babu \& Singh}{Babu \& Singh}{1983}]{babu1983}
Babu G.~J.,  Singh K.,  1983, The Annals of Statistics, pp 999--1003

\bibitem[\protect\citeauthoryear{{Beifiori}, {Courteau}, {Corsini}  \&
  {Zhu}}{{Beifiori} et~al.}{2012}]{beifiori2012}
{Beifiori} A.,  {Courteau} S.,  {Corsini} E.~M.,   {Zhu} Y.,  2012, \mn@doi
  [\mnras] {10.1111/j.1365-2966.2011.19903.x}, \href
  {http://adsabs.harvard.edu/abs/2012MNRAS.419.2497B} {419, 2497}

\bibitem[\protect\citeauthoryear{Boggs, Byrd  \& Schnabel}{Boggs
  et~al.}{1987}]{boggs1987}
Boggs P.~T.,  Byrd R.~H.,   Schnabel R.~B.,  1987, SIAM Journal on Scientific
  and Statistical Computing, 8, 1052

\bibitem[\protect\citeauthoryear{Boggs, Spiegelman, Donaldson  \&
  Schnabel}{Boggs et~al.}{1988}]{boggs1988}
Boggs P.~T.,  Spiegelman C.~H.,  Donaldson J.~R.,   Schnabel R.~B.,  1988,
  \mn@doi [Journal of Econometrics]
  {http://dx.doi.org/10.1016/0304-4076(88)90032-2}, 38, 169

\bibitem[\protect\citeauthoryear{{Booth} \& {Schaye}}{{Booth} \&
  {Schaye}}{2009}]{booth2009}
{Booth} C.~M.,  {Schaye} J.,  2009, \mn@doi [\mnras]
  {10.1111/j.1365-2966.2009.15043.x}, \href
  {http://adsabs.harvard.edu/abs/2009MNRAS.398...53B} {398, 53}

\bibitem[\protect\citeauthoryear{Broemeling \& Broemeling}{Broemeling \&
  Broemeling}{2003}]{broemeling2003}
Broemeling L.,  Broemeling A.,  2003, \mn@doi [Biometrika]
  {10.1093/biomet/90.3.728}, 90, 728

\bibitem[\protect\citeauthoryear{Calin \& Udriste}{Calin \&
  Udriste}{2014}]{calin2014}
Calin O.,  Udriste C.,  2014, Geometric Modeling in Probability and Statistics.
Mathematics and Statistics, Springer International Publishing

\bibitem[\protect\citeauthoryear{{Choi}, {Naab}, {Ostriker}, {Johansson}  \&
  {Moster}}{{Choi} et~al.}{2014}]{choi2014}
{Choi} E.,  {Naab} T.,  {Ostriker} J.~P.,  {Johansson} P.~H.,   {Moster} B.~P.,
   2014, \mn@doi [\mnras] {10.1093/mnras/stu874}, \href
  {http://adsabs.harvard.edu/abs/2014MNRAS.442..440C} {442, 440}

\bibitem[\protect\citeauthoryear{Crane, Weischedel  \& Wardetzky}{Crane
  et~al.}{2013}]{crane2013}
Crane K.,  Weischedel C.,   Wardetzky M.,  2013, ACM Trans. Graph., 32

\bibitem[\protect\citeauthoryear{{Di Matteo}, {Springel}  \& {Hernquist}}{{Di
  Matteo} et~al.}{2005}]{dimatteo2005}
{Di Matteo} T.,  {Springel} V.,   {Hernquist} L.,  2005, \mn@doi [\nat]
  {10.1038/nature03335}, \href
  {http://adsabs.harvard.edu/abs/2005Natur.433..604D} {433, 604}

\bibitem[\protect\citeauthoryear{{Di Matteo}, {Colberg}, {Springel},
  {Hernquist}  \& {Sijacki}}{{Di Matteo} et~al.}{2008}]{dimatteo2008}
{Di Matteo} T.,  {Colberg} J.,  {Springel} V.,  {Hernquist} L.,   {Sijacki} D.,
   2008, \mn@doi [\apj] {10.1086/524921}, \href
  {http://adsabs.harvard.edu/abs/2008ApJ...676...33D} {676, 33}

\bibitem[\protect\citeauthoryear{{Feigelson} \& {Babu}}{{Feigelson} \&
  {Babu}}{1992}]{feigelson1992}
{Feigelson} E.~D.,  {Babu} G.~J.,  1992, \mn@doi [\apj] {10.1086/171766}, \href
  {http://adsabs.harvard.edu/abs/1992ApJ...397...55F} {397, 55}

\bibitem[\protect\citeauthoryear{Feigelson \& Babu}{Feigelson \&
  Babu}{2012}]{feigelson2012}
Feigelson E.,  Babu G.,  2012, Modern Statistical Methods for Astronomy: With R
  Applications.
Cambridge University Press

\bibitem[\protect\citeauthoryear{{Ferrarese} \& {Merritt}}{{Ferrarese} \&
  {Merritt}}{2000}]{ferrarese2000}
{Ferrarese} L.,  {Merritt} D.,  2000, \mn@doi [\apjl] {10.1086/312838}, \href
  {http://adsabs.harvard.edu/abs/2000ApJ...539L...9F} {539, L9}

\bibitem[\protect\citeauthoryear{{Foreman-Mackey}, {Hogg}, {Lang}  \&
  {Goodman}}{{Foreman-Mackey} et~al.}{2013}]{emcee2013}
{Foreman-Mackey} D.,  {Hogg} D.~W.,  {Lang} D.,   {Goodman} J.,  2013, \mn@doi
  [\pasp] {10.1086/670067}, \href
  {http://adsabs.harvard.edu/abs/2013PASP..125..306F} {125, 306}

\bibitem[\protect\citeauthoryear{Fraser, Reid, Marras  \& Yi}{Fraser
  et~al.}{2010}]{fraser2010}
Fraser D. A.~S.,  Reid N.,  Marras E.,   Yi G.~Y.,  2010, \mn@doi [Journal of
  the Royal Statistical Society: Series B (Statistical Methodology)]
  {10.1111/j.1467-9868.2010.00750.x}, 72, 631

\bibitem[\protect\citeauthoryear{Fujiki \& Akaho}{Fujiki \&
  Akaho}{2009}]{fujiki2009}
Fujiki J.,  Akaho S.,  2009, in 2009 IEEE 12th International Conference on
  Computer Vision Workshops, ICCV Workshops. pp 250--255,
  \mn@doi{10.1109/ICCVW.2009.5457693}

\bibitem[\protect\citeauthoryear{{Gebhardt} et~al.,}{{Gebhardt}
  et~al.}{2000}]{gebhardt2000}
{Gebhardt} K.,  et~al., 2000, \mn@doi [\apjl] {10.1086/312840}, \href
  {http://adsabs.harvard.edu/abs/2000ApJ...539L..13G} {539, L13}

\bibitem[\protect\citeauthoryear{Gelman, Carlin, Stern, Dunson, Vehtari  \&
  Rubin}{Gelman et~al.}{2013}]{gelman2013}
Gelman A.,  Carlin J.,  Stern H.,  Dunson D.,  Vehtari A.,   Rubin D.,  2013,
  Bayesian Data Analysis, Third Edition.
Chapman \& Hall/CRC Texts in Statistical Science, Taylor \& Francis

\bibitem[\protect\citeauthoryear{George \& McCulloch}{George \&
  McCulloch}{1993}]{george1993}
George E.~I.,  McCulloch R.,  1993, \mn@doi [Journal of Statistical Planning
  and Inference] {http://dx.doi.org/10.1016/0378-3758(93)90086-L}, 37, 169

\bibitem[\protect\citeauthoryear{{Graham}}{{Graham}}{2016}]{graham2016}
{Graham} A.~W.,  2016, \mn@doi [Galactic Bulges]
  {10.1007/978-3-319-19378-6_11}, \href
  {http://adsabs.harvard.edu/abs/2016ASSL..418..263G} {418, 263}

\bibitem[\protect\citeauthoryear{{Graham}, {Onken}, {Athanassoula}  \&
  {Combes}}{{Graham} et~al.}{2011}]{graham2011}
{Graham} A.~W.,  {Onken} C.~A.,  {Athanassoula} E.,   {Combes} F.,  2011,
  \mn@doi [\mnras] {10.1111/j.1365-2966.2010.18045.x}, \href
  {http://adsabs.harvard.edu/abs/2011MNRAS.412.2211G} {412, 2211}

\bibitem[\protect\citeauthoryear{Gull}{Gull}{1989}]{gull1989}
Gull S.~F.,  1989, in , {Maximum Entropy and Bayesian Methods}.
Springer, pp 511--518

\bibitem[\protect\citeauthoryear{{G{\"u}ltekin} et~al.,}{{G{\"u}ltekin}
  et~al.}{2009}]{gultekin2009}
{G{\"u}ltekin} K.,  et~al., 2009, \mn@doi [\apj] {10.1088/0004-637X/698/1/198},
  \href {http://adsabs.harvard.edu/abs/2009ApJ...698..198G} {698, 198}

\bibitem[\protect\citeauthoryear{{Hogg}, {Bovy}  \& {Lang}}{{Hogg}
  et~al.}{2010}]{hogg2010}
{Hogg} D.~W.,  {Bovy} J.,   {Lang} D.,  2010, preprint, \href
  {http://adsabs.harvard.edu/abs/2010arXiv1008.4686H} {} (\mn@eprint {arXiv}
  {1008.4686})

\bibitem[\protect\citeauthoryear{Hunter}{Hunter}{2007}]{matplotlib}
Hunter J.~D.,  2007, \mn@doi [Computing In Science \& Engineering]
  {10.1109/MCSE.2007.55}, 9, 90

\bibitem[\protect\citeauthoryear{{Isobe}, {Feigelson}, {Akritas}  \&
  {Babu}}{{Isobe} et~al.}{1990}]{isobe1990}
{Isobe} T.,  {Feigelson} E.~D.,  {Akritas} M.~G.,   {Babu} G.~J.,  1990,
  \mn@doi [\apj] {10.1086/169390}, \href
  {http://adsabs.harvard.edu/abs/1990ApJ...364..104I} {364, 104}

\bibitem[\protect\citeauthoryear{Jaynes}{Jaynes}{1983}]{jaynes1983}
Jaynes E.,  1983, in Rosenkrantz R.~D.,  ed., , Vol.~158, {E.T.~Jaynes: Papers
  on probability, statistics and statistical physics}.
D.~Reidel Publishing Company, Dordrecht, Holland, pp 190--209

\bibitem[\protect\citeauthoryear{Jaynes}{Jaynes}{1991}]{jaynes1991}
Jaynes E.~T.,  1991, {Straight Line Fitting -- A Bayesian Solution},
  Unpublished, available online at \url{http://bayes.wustl.edu/sfg/line.pdf}

\bibitem[\protect\citeauthoryear{Jaynes}{Jaynes}{2003}]{jaynes2003}
Jaynes E.~T.,  2003, {Probability Theory - The Logic of Science}.
Cambridge University Press, Cambridge, United Kingdom

\bibitem[\protect\citeauthoryear{{Johansson}, {Naab}  \& {Burkert}}{{Johansson}
  et~al.}{2009a}]{johansson2009a}
{Johansson} P.~H.,  {Naab} T.,   {Burkert} A.,  2009a, \mn@doi [\apj]
  {10.1088/0004-637X/690/1/802}, \href
  {http://adsabs.harvard.edu/abs/2009ApJ...690..802J} {690, 802}

\bibitem[\protect\citeauthoryear{{Johansson}, {Burkert}  \& {Naab}}{{Johansson}
  et~al.}{2009b}]{johansson2009b}
{Johansson} P.~H.,  {Burkert} A.,   {Naab} T.,  2009b, \mn@doi [\apjl]
  {10.1088/0004-637X/707/2/L184}, \href
  {http://adsabs.harvard.edu/abs/2009ApJ...707L.184J} {707, L184}

\bibitem[\protect\citeauthoryear{Jones, Oliphant, Peterson  et~al.}{Jones
  et~al.}{01  }]{scipy}
Jones E.,  Oliphant T.,  Peterson P.,   et~al., 2001--, {SciPy}: Open source
  scientific tools for {Python}, \url {http://www.scipy.org/}

\bibitem[\protect\citeauthoryear{Jupp \& Kent}{Jupp \& Kent}{1987}]{jupp1987}
Jupp P.~E.,  Kent J.~T.,  1987, Journal of the Royal Statistical Society.
  Series C (Applied Statistics), 36, 34

\bibitem[\protect\citeauthoryear{{Kelly}}{{Kelly}}{2007}]{kelly2007}
{Kelly} B.~C.,  2007, \mn@doi [\apj] {10.1086/519947}, \href
  {http://adsabs.harvard.edu/abs/2007ApJ...665.1489K} {665, 1489}

\bibitem[\protect\citeauthoryear{Lee}{Lee}{2013}]{lee2013}
Lee J.,  2013, Introduction to Smooth Manifolds.
Graduate Texts in Mathematics, Springer New York

\bibitem[\protect\citeauthoryear{{Magorrian} et~al.,}{{Magorrian}
  et~al.}{1998}]{magorrian1998}
{Magorrian} J.,  et~al., 1998, \mn@doi [\aj] {10.1086/300353}, \href
  {http://adsabs.harvard.edu/abs/1998AJ....115.2285M} {115, 2285}

\bibitem[\protect\citeauthoryear{{Markwardt}}{{Markwardt}}{2009}]{markwardt2009}
{Markwardt} C.~B.,  2009, in {Bohlender} D.~A.,  {Durand} D.,   {Dowler} P.,
  eds,  Astronomical Society of the Pacific Conference Series Vol. 411,
  Astronomical Data Analysis Software and Systems XVIII. p.~251 (\mn@eprint
  {arXiv} {0902.2850})

\bibitem[\protect\citeauthoryear{{McConnell} \& {Ma}}{{McConnell} \&
  {Ma}}{2013}]{mcconnell2013}
{McConnell} N.~J.,  {Ma} C.-P.,  2013, \mn@doi [\apj]
  {10.1088/0004-637X/764/2/184}, \href
  {http://adsabs.harvard.edu/abs/2013ApJ...764..184M} {764, 184}

\bibitem[\protect\citeauthoryear{{McConnell}, {Ma}, {Gebhardt}, {Wright},
  {Murphy}, {Lauer}, {Graham}  \& {Richstone}}{{McConnell}
  et~al.}{2011}]{mcconnell2011}
{McConnell} N.~J.,  {Ma} C.-P.,  {Gebhardt} K.,  {Wright} S.~A.,  {Murphy}
  J.~D.,  {Lauer} T.~R.,  {Graham} J.~R.,   {Richstone} D.~O.,  2011, \mn@doi
  [\nat] {10.1038/nature10636}, \href
  {http://adsabs.harvard.edu/abs/2011Natur.480..215M} {480, 215}

\bibitem[\protect\citeauthoryear{Meurer et~al.,}{Meurer et~al.}{2017}]{sympy}
Meurer A.,  et~al., 2017, \mn@doi [PeerJ Computer Science]
  {10.7717/peerj-cs.103}, 3, e103

\bibitem[\protect\citeauthoryear{{Novak}, {Faber}  \& {Dekel}}{{Novak}
  et~al.}{2006}]{novak2006}
{Novak} G.~S.,  {Faber} S.~M.,   {Dekel} A.,  2006, \mn@doi [\apj]
  {10.1086/498333}, \href {http://adsabs.harvard.edu/abs/2006ApJ...637...96N}
  {637, 96}

\bibitem[\protect\citeauthoryear{{Park}, {Kelly}, {Woo}  \& {Treu}}{{Park}
  et~al.}{2012}]{park2012}
{Park} D.,  {Kelly} B.~C.,  {Woo} J.-H.,   {Treu} T.,  2012, \mn@doi [\apjs]
  {10.1088/0067-0049/203/1/6}, \href
  {http://adsabs.harvard.edu/abs/2012ApJS..203....6P} {203, 6}

\bibitem[\protect\citeauthoryear{Pearson}{Pearson}{1901}]{pearson1901}
Pearson K.,  1901, {Philosophical Magazine Series 6}, 2, 559

\bibitem[\protect\citeauthoryear{Pennec}{Pennec}{2006}]{pennec2006}
Pennec X.,  2006, \mn@doi [J. Math. Imaging Vis.] {10.1007/s10851-006-6228-4},
  25, 127

\bibitem[\protect\citeauthoryear{{Press}, {Teukolsky}, {Vetterling}  \&
  {Flannery}}{{Press} et~al.}{1992}]{press1992}
{Press} W.~H.,  {Teukolsky} S.~A.,  {Vetterling} W.~T.,   {Flannery} B.~P.,
  1992, {Numerical recipes in FORTRAN. The art of scientific computing}.
Cambridge: University Press, c1992, 2nd ed.

\bibitem[\protect\citeauthoryear{{Robotham} \& {Obreschkow}}{{Robotham} \&
  {Obreschkow}}{2015}]{robotham2015}
{Robotham} A.~S.~G.,  {Obreschkow} D.,  2015, \mn@doi [\pasa]
  {10.1017/pasa.2015.33}, \href
  {http://adsabs.harvard.edu/abs/2015PASA...32...33R} {32, e033}

\bibitem[\protect\citeauthoryear{{Saglia} et~al.,}{{Saglia}
  et~al.}{2016}]{saglia2016}
{Saglia} R.~P.,  et~al., 2016, \mn@doi [\apj] {10.3847/0004-637X/818/1/47},
  \href {http://adsabs.harvard.edu/abs/2016ApJ...818...47S} {818, 47}

\bibitem[\protect\citeauthoryear{{Sijacki}, {Springel}, {Di Matteo}  \&
  {Hernquist}}{{Sijacki} et~al.}{2007}]{sijacki2007}
{Sijacki} D.,  {Springel} V.,  {Di Matteo} T.,   {Hernquist} L.,  2007, \mn@doi
  [\mnras] {10.1111/j.1365-2966.2007.12153.x}, \href
  {http://adsabs.harvard.edu/abs/2007MNRAS.380..877S} {380, 877}

\bibitem[\protect\citeauthoryear{{Sijacki}, {Vogelsberger}, {Genel},
  {Springel}, {Torrey}, {Snyder}, {Nelson}  \& {Hernquist}}{{Sijacki}
  et~al.}{2015}]{sijacki2015}
{Sijacki} D.,  {Vogelsberger} M.,  {Genel} S.,  {Springel} V.,  {Torrey} P.,
  {Snyder} G.~F.,  {Nelson} D.,   {Hernquist} L.,  2015, \mn@doi [\mnras]
  {10.1093/mnras/stv1340}, \href
  {http://adsabs.harvard.edu/abs/2015MNRAS.452..575S} {452, 575}

\bibitem[\protect\citeauthoryear{Stigler}{Stigler}{1981}]{stigler1981}
Stigler S.~M.,  1981, \mn@doi [Ann. Statist.] {10.1214/aos/1176345451}, 9, 465

\bibitem[\protect\citeauthoryear{{Tremaine} et~al.,}{{Tremaine}
  et~al.}{2002}]{tremaine2002}
{Tremaine} S.,  et~al., 2002, \mn@doi [\apj] {10.1086/341002}, \href
  {http://adsabs.harvard.edu/abs/2002ApJ...574..740T} {574, 740}

\bibitem[\protect\citeauthoryear{{Williams}, {Bureau}  \&
  {Cappellari}}{{Williams} et~al.}{2010}]{williams2010}
{Williams} M.~J.,  {Bureau} M.,   {Cappellari} M.,  2010, \mn@doi [\mnras]
  {10.1111/j.1365-2966.2010.17406.x}, \href
  {http://adsabs.harvard.edu/abs/2010MNRAS.409.1330W} {409, 1330}

\bibitem[\protect\citeauthoryear{Wolf \& Zierau}{Wolf \&
  Zierau}{1996}]{wolf1996}
Wolf J.~A.,  Zierau R.,  1996, Riemannian exponential maps and decompositions
  of reductive Lie groups.
Springer, pp 349--354

\bibitem[\protect\citeauthoryear{Zellner}{Zellner}{1971}]{zellner1971}
Zellner A.,  1971, An introduction to Bayesian inference in econometrics.
Wiley series in probability and mathematical statistics: Applied probability
  and statistics, J. Wiley

\bibitem[\protect\citeauthoryear{{van den Bosch}}{{van den
  Bosch}}{2016}]{bosch2016}
{van den Bosch} R.~C.~E.,  2016, \mn@doi [\apj] {10.3847/0004-637X/831/2/134},
  \href {http://adsabs.harvard.edu/abs/2016ApJ...831..134V} {831, 134}

\bibitem[\protect\citeauthoryear{van~der Walt, Colbert  \& Varoquaux}{van~der
  Walt et~al.}{2011}]{numpy}
van~der Walt S.,  Colbert S.~C.,   Varoquaux G.,  2011, \mn@doi [Computing in
  Science Engineering] {10.1109/MCSE.2011.37}, 13, 22

\makeatother
\end{thebibliography}




\appendix

\section{Derivation of the posterior distribution}\label{sc:appendix-a}

We present a derivation of the
likelihood, equation~\eqref{eq:likelihood}, and the posterior
distribution, equation~\eqref{eq:postdist}, following the Bayesian style
promoted in \citet{jaynes2003}.
We consider the following propositions:
\begin{itemize}
        \item $\eta$ = `the true value of the observable is $\eta\in M$'
        \item $X$ = `a value $x\in M$ was measured for the observable'
        \item $V$ = `a detection was made'
        \item $H$ = `the true value of the observable is drawn from the
            distribution $\intdist$ defined on a relation $S$, with parameters
            $\intpars$ and $\relpars$, respectively'
\end{itemize}
In addition, we will use $I$ to specify all the other relevant
prior information. This includes the prior distribution
$\pi_{\relpars,\intpars}$ of $\relpars$ and
$\intpars$, the measurement error distribution $h(\eta;x)$ and
the fact that the probability of a detection for a
value $\eta$ of the observable is given by $\detdist(\eta)$.

At the outset, we then know the following probabilities
\begin{align}
    P(\eta|HI) &= \intdist(\eta;\relpars,\intpars) := \intdist(\para(\eta;\relpars),\ortho(\eta;\relpars);\intpars) \\
    P(\eta|XI) &= h(\eta;x) \\
    P(V|\eta I) &= \detdist(\eta).
\end{align}
We will need the probability $P(X|\eta VI)$ of a measured value, given a true value and
the fact that there is a detection. This is obtained with Bayes' theorem
\begin{equation}
    P(X|\eta VI) = P(\eta|XVI) \frac{P(X|VI)}{P(\eta|VI)} = h(\eta;x),
\end{equation}
as a function of $x$ (which we write as $h(x;\eta)$ in the following), 
since the prior probabilities $P(X|VI)$ and
$P(\eta|VI)$ must be uninformative and equal everywhere, since nothing
in $I$ tells us where the true and measured values are a priori. 

We can now compute the probability of a measured value given the true
value, yielding
\begin{equation}
\begin{split}
    P(X|\eta I) 
    &= P(X(V+\bar{V})|\eta I) = P(XV|\eta I) + P(X\bar{V}|\eta I) \\
    &= P(X|\eta VI)P(V|\eta I) + P(X|\eta \bar{V} I) P(\bar{V}|\eta I) \\
    &= h(x;\eta)\detdist(\eta) + 0\cdot\left[ 1-\detdist(\eta) \right] \\
    &= h(x;\eta)\detdist(\eta),
\end{split}
\end{equation}
where $\bar{V}$ is the negation of $V$ (i.e.\ there was no detection).
Here we used the fact that $V$ and $\bar{V}$ form a complete
($P(V+\bar{V}|I)=1$) and independent set of propositions, together with the identity
$P(AB|C) = P(A|C)P(B|AC) = P(B|C)P(A|BC)$.

Using Bayes' theorem again we can now obtain the probability $P(H|XI)$ that
the data obey the relation and are drawn from $\intdist$, given the
measured value. The theorem gives 
\begin{equation}\label{eq:h-bayes}
    P(H|XI) = P(X|HI)\frac{P(H|I)}{P(X|I)}.
\end{equation}
Since the $\eta$ also form a complete and independent
set of propositions (the true value must be somewhere, and the possible
positions are independent), we can write
\begin{equation}
\begin{split}
    P(X|HI) &= \int P(X\eta|HI)\ud\eta = \int P(X|\eta HI)P(\eta|HI)\ud\eta \\
            &= \int h(x;\eta)\detdist(\eta) \intdist(\eta;\relpars,\intpars) \ud\eta  \\
            &= \mathcal{L}(\relpars,\intpars|x),
\end{split}
\end{equation}
which gives the likelihood, equation~\eqref{eq:likelihood}.
The factor $P(H|I) = \pi_{\relpars,\intpars}(\relpars,\intpars)$ is the parameter prior probability,
and the denominator is a normalizing constant, given formally by
\begin{equation}
\begin{split}
    P(X|I) &= \int P(XH|I)\ud H = \int P(X|HI) P(H|I) \ud H \\
           &= \iint h(x;\eta)\detdist(\eta)\intdist(\eta;\relpars,\intpars)
    \pi_{\relpars,\intpars}(\relpars,\intpars) \ud\eta\, \ud\relpars\ud\intpars.
\end{split}
\end{equation}
We then have the posterior probability, equation~\eqref{eq:postdist},
\begin{equation}
    P(H|XI) = \frac{%
        \int h(x;\eta)\detdist(\eta) \intdist(\eta;\relpars,\intpars) \ud\eta
        \,\,\pi_{\relpars,\intpars}(\relpars,\intpars)
    }{%
        \iint h(x;\eta)\detdist(\eta)\intdist(\eta;\relpars,\intpars)
        \pi_{\relpars,\intpars}(\relpars,\intpars) \ud\eta\, \ud\relpars\ud\intpars
    }.
\end{equation}

\section{An example of limit distributions}\label{sc:appendix-b}

\add{Here we present a simple example of how the framework presented in the
paper works with upper limits, and how analytic results can be obtained
for upper limits as well. This is desirable since typically the analytic
expression for the likelihood is much faster to compute than performing
a numerical integration to obtain the required value.
As mentioned in Section~\ref{sc:intscat}, limits are taken into account
by including them into the measurement error distribution.  This can be
done for the examples in Sections~\ref{sc:nd-case}--\ref{sc:lincase} by
substituting one or more degrees of freedom $x_i$
in equations \eqref{eq:error-nd-gaussian} and \eqref{eq:2dgaussian}
with suitable one-dimensional distributions. Conceptually the simplest
possibility is the uniform distribution, so that
\begin{equation}\label{eq:upperlimit}
    u(x_i;x_i^u)= \frac{\chi_{[0,x_i^u]}(x_i)}{x_i^u}, 
\end{equation}
where $\chi_{[a,b]}(x)=1$ if $x\in[a,b]$ and
$0$ otherwise, and $x_i^u$ is the limiting value. Other choices with
less pronounced cutoffs are naturally also possible.
If the measurements are already given in logarithmic units with base
$k$, the form
\begin{equation}\label{eq:logupperlimit}
    u_{\log_k}(x_i;x_i^u)=u(k^{x_i};k^{x_i^u})k^{x_i}\log k 
\end{equation}
must be used instead. Lower limits can be introduced in an analogous
manner.}

\add{Analytic results of reasonable complexity based on these limit
distributions can be obtained for the example in
Section~\ref{sc:lincase}. For the linear case we have
\begin{equation}
\label{eq:upperx}
\begin{split}
    &\mathcal{L}_x(\alpha,\beta,\sigma|x^u,y_0,\sigma_y) = \\
    &\quad \frac{1}{2x^u}\frac{\sqrt{1+\beta^2}}{\beta}
    \left\{
        \erf\left(
        \frac{\alpha + x^u\beta - y_0}{\sqrt{2(\sigma_y^2 + (1+\beta^2)\sigma^2)}}
        \right)\right.
        \\
        &\quad\quad\quad\quad
        \left.
        -\erf\left(
        \frac{\alpha - y_0}{\sqrt{2(\sigma_y^2 + (1+\beta^2)\sigma^2)}}
        \right)
    \right\} 
\end{split} 
\end{equation}
\begin{equation}
\label{eq:uppery}
\begin{split}
    &\mathcal{L}_y(\alpha,\beta,\sigma|x_0,y^u,\sigma_x) = \\
    &\quad\frac{1}{2y^u}\sqrt{1+\beta^2}
    \left\{
        \erf\left(
        \frac{\alpha + x_0\beta}{\sqrt{2(\beta^2\sigma_x^2 + (1+\beta^2)\sigma^2)}}
        \right)\right.
        \\
        &\quad\quad\quad\quad
        \left.
        -\erf\left(
        \frac{\alpha + x_0\beta - y^u}{\sqrt{2(\beta^2\sigma_x^2 + (1+\beta^2)\sigma^2)}}
        \right)
    \right\},
\end{split}
\end{equation}
for the upper limits in $x$ and $y$ directions respectively, where $x^u$
and $y^u$ are the corresponding limiting values. If the measurements
have been given in logarithmic scale with base $k$, we have instead
\begin{equation}
\label{eq:logupperx}
\begin{split}
    &\mathcal{L}_{\log_k x}(\alpha,\beta,\sigma|x^u,y_0,\sigma_y) = \\ 
    &\, \frac{\log k}{2 x^u}\sqrt{\frac{1+\beta^2}{\beta^2}}
    k^{ \frac{2(y_0-\alpha)\beta + [\sigma_y^2 + (1+\beta^2)\sigma^2]\log k}%
    {2\beta^2} } 
    \\
    &\,
    \erfc\left(
    \frac{
        (y_0-\alpha)\beta+[\sigma_y^2+(1+\beta^2)\sigma^2]\log
        k-\beta^2\log_k x^u
    }{
        \sqrt{2[\sigma_y^2+(1+\beta^2)\sigma^2]}\abs{\beta}
    }
    \right)
\end{split}
\end{equation}
\begin{equation}
\label{eq:loguppery}
\begin{split}
    &\mathcal{L}_{\log_k y}(\alpha,\beta,\sigma|x_0,y^u,\sigma_x) = \\
    &\, \frac{\log k}{2 y^u}\sqrt{1+\beta^2}
    k^{ \alpha + x_0\beta +
    \frac{1}{2}[\beta^2\sigma_x^2+(1+\beta^2)\sigma^2]\log k } 
    \\
    &\,
    \erfc\left(
    \frac{
        \alpha + x_0\beta + [\beta^2\sigma_x^2+(1+\beta^2)\sigma^2]\log k
        - \log_k y^u
    }{
        \sqrt{2[\beta^2\sigma_x^2+(1+\beta^2)\sigma^2]}
    }
    \right),
\end{split}
\end{equation}
where $\erfc(x)=1-\erf(x)$. Note that here the limiting values $x^u$ and
$y^u$ are given in linear scale.
For numerical applications, it should be noted that the arguments of the $k$-exponential and the
$\erfc$ function may have large numerical values, and asymptotic expansions
should be used when necessary.}


\bsp	
\label{lastpage}
\end{document}